\documentclass[a4paper,12pt]{article}
\usepackage{amsmath,enumerate}
\usepackage{graphicx}
\usepackage{multirow}
\usepackage{xcolor}
\usepackage{minitoc}
\usepackage{braket}
\usepackage{subfig}
\usepackage{slashed}
\usepackage{float}
\usepackage{xcolor}
\usepackage{cite}
\usepackage{datetime}
\definecolor{MyBlue}{rgb}{0.25,0.5,0.75}
\usepackage[colorlinks=true,allcolors=blue]{hyperref}
\newcommand{\lapprox}{%
	\mathrel{%
		\setbox0=\hbox{$<$}
		\raise0.6ex\copy0\kern-\wd0
		\lower0.65ex\hbox{$\sim$}
}}
\newcommand{\gapprox}{%
	\mathrel{%
		\setbox0=\hbox{$>$}
		\raise0.6ex\copy0\kern-\wd0
		\lower0.65ex\hbox{$\sim$}
}}
\newdateformat{monthyeardate}{%
	\monthname[\THEMONTH], \THEYEAR}
\begin{document}

\begin{center}

{\Large \bf Lepton flavor violation in the Majorana and Dirac scotogenic models}\\[20mm]

Raghavendra Srikanth Hundi\\
Department of Physics, Indian Institute of Technology Hyderabad,\\
Kandi - 502 284, India.\\[5mm]
E-mail: rshundi@phy.iith.ac.in \\[20mm]

\end{center}

\begin{abstract}

In this work we have considered two minimal versions of scotogenic models, where
neutrinos acquire masses through a radiative mechanism. We call these two models
as Majorana and Dirac scotogenic models. In the former model, neutrinos have
Majorana nature, and in the later one, neutrinos are Dirac particles. These two
models are related to each other in terms of additional fields and symmetries of
the model. Hence, to compare these two models in future experiments, we have analyzed
lepton flavor violating (LFV) processes in both of them, in the charged lepton sector.
We have found that the 3-body LFV decays in both these models can get different
contributions. Among all the LFV decays and after satisfying relevant constraints,
we have found that $\tau\to3\mu$ can have a branching ratio as high as
$10^{-10}(10^{-11})$ in the Majorana(Dirac) scotogenic model.

\end{abstract}
\newpage

\section{Introduction}\label{s1}

From the neutrino oscillation experiments \cite{Kajita:2016cak,McDonald:2016ixn}
and cosmological data \cite{Planck:2018vyg}, it is known that
neutrinos have non-zero and tiny masses. Non-zero masses for neutrinos can be conceived
in the standard model by proposing singlet right-handed neutrinos. However, smallness
of their masses can be explained only by unnaturally suppressing the Yukawa couplings.
As a result of this, to understand the origin of neutrino masses, the standard model
should be expanded with new fields and symmetries \cite{Quigg:2004is,Ellis:2009pz}.
Several models have been proposed on the origin of neutrino masses. For instance,
see \cite{King:2003jb,Altarelli:2006ri} for a review on these models.
Most of these models are based on seesaw mechanism \cite{Minkowski:1977sc,
Gell-Mann:1979vob,Mohapatra:1979ia,
Schechter:1980gr,Magg:1980ut,Mohapatra:1980yp,Lazarides:1980nt,Foot:1988aq},
according to which, conservation of lepton number is violated, and hence, neutrinos
are Majorana particles. Whether lepton number is violated or not is not yet known
in experiments. As a result of this, models which conserve lepton number are also
proposed, according to which neutrinos are Dirac particles.
See \cite{Mohapatra:1987hh,Chang:1986bp,Mohapatra:1987nx,Balakrishna:1988bn,
Ma:1989tz,Babu:1988yq,Branco:1987yg,Babu:1989fg,Nasri:2001ax} for an early literature
of models on Dirac neutrino masses. From the current status
of experiments, neutrinos can have either Majorana or Dirac nature and it is still
an open problem to know which of these two possibilities is true.

To understand the origin of neutrino masses, a particular class of models, known as
scotogenic models are studied. In these models, neutrinos acquire masses through a loop
induced mechanism, and as a result of that, loop suppression factor gives a natural
suppression for explaining the smallness of neutrino masses. Scotogenic models
can be classified into two categories, which depend on if the neutrinos in these models
have Majorana or Dirac nature. In the category of scotogenic models, where neutrinos
are Majorana particles, there exist extensive literature. In this category,
see \cite{Ma:2006km,Ma:2006uv,Ma:2007yx,Ma:2007gq,Ma:2007kt,Ma:2008ba,Gu:2008zf,Ma:2008cu},
\cite{Vicente:2014wga,Molinaro:2014lfa,Lindner:2016kqk,Escribano:2020iqq,Sarazin:2021nwo,
Portillo-Sanchez:2023kbz,Alvarez:2023dzz,
Abada:2023znk,Ahriche:2023hho,
Borah:2023hqw,Avnish:2023wxk,Singh:2023eye,Kitabayashi:2023rje,Karan:2023adm,
Herms:2023cyy,Ismael:2023jlp,Escribano:2023hxj} and \cite{Hundi:2022iva,Hundi:2015kea}
for early models, some recent phenomenological works and our own works, respectively.
On the other hand, in the category of scotogenic models, where neutrinos are Dirac
particles, see \cite{Farzan:2012sa,Ma:2016mwh,Wang:2017mcy,Yao:2017vtm,Calle:2018ovc,
Ma:2019yfo,Ma:2019iwj,Ma:2019byo,CentellesChulia:2019xky,Jana:2019mgj,Leite:2020wjl,
Guo:2020qin,Bernal:2021ezl,De:2021crr,Mishra:2021ilq,Chowdhury:2022jde,Li:2022chc,
Hazarika:2022tlc,Borah:2022enh,Hundi:2023tdq} for various models and phenomenological
works. Notice that, apart scotogenic models, scoto-seesaw mechanism
\cite{Rojas:2018wym} is proposed,
according to which one neutrino acquire mass at tree level and another neutrino
acquire mass through a radiative process. See \cite{Kumar:2023moh,Kumar:2024zfb}
for recent works based on this.
However, in this work, our interest is on scotogenic models.

Since the scotogenic models
are viable in explaining the smallness of neutrino masses, in this work, we consider
two minimal versions of these models, which are named as Majorana and Dirac scotogenic
models. In the Majorana scotogenic model (MSM) \cite{Ma:2006km},
neutrinos are Majorana particles.
Whereas in the Dirac scotogenic model (DSM) \cite{Farzan:2012sa}, neutrinos are purely
Dirac particles. Both MSM and DSM are extensions to the standard model with new fields
and symmetries. In the MSM, the additional fields are the scalar doublet $\eta^M$ and the
singlet Majorana fermions $N^M_k$, $k=1,2,3$. To achieve loop induced masses to neutrinos
in the MSM \cite{Ma:2006km}, a discrete symmetry $Z_2$ is introduced, under which
$\eta^M$ and $N^M_k$ are
odd and all the standard model fields are even. On the other hand, in the DSM
\cite{Farzan:2012sa}, the
additional fields are the scalar doublet $\eta^D$, the singlet scalar $\chi$ and the
following singlet Weyl fermions: $\nu^c_\alpha$, $N_k$, $N^c_k$. Here, $\alpha=e,\mu,\tau$.
The fields $\nu^c_\alpha$ combine with the left-handed neutrinos of lepton doublets
to form Dirac neutrinos and $N_k$, $N^c_k$ combine together to form Dirac fermions
$N^D_k$ in the DSM. To generate radiative masses to Dirac neutrinos in this model
\cite{Farzan:2012sa},
a discrete symmetry $Z_2^{(A)}\times Z_2^{(B)}$ is proposed, under which all the above
described fields are charged and the standard model fields are even. From the description
given above for both the MSM and DSM, we see that there is a similarity between
these two models in terms of field content and symmetries. In fact, note that,
we have used the suffixes $M$ and
$D$ to distinguish similar fields in the MSM and DSM. Since these models are
similar to each other, it is worth to compare observable quantities of these two models,
which is one of the aims of the current work.

As described above, the MSM and DSM can address the problem of origin of neutrino masses.
The other problem that needs to be addressed in these models is related to the
neutrino mixing angles, which is described below.
From the oscillation data \cite{deSalas:2020pgw}, it is known that there exists
two large mixing angles
and one small mixing angle in the neutrino sector. A consequence of these mixing
angles is that the mass matrix for neutrinos in the flavor basis will have nonzero
off-diagonal elements. Such a structure of mass matrix can be explained in the MSM
and DSM, only if the Yukawa couplings of the lepton doublets have off-diagonal
elements. We demonstrate this statement explicitly in the next section.
Generally, the off-diagonal elements of Yukawa couplings drive lepton flavor violating
(LFV) processes such as $\ell_\alpha\to\ell_\beta\gamma$,
$\ell_\alpha\to\ell_\beta\ell_\rho\overline{\ell}_\delta$ and $\mu N\to eN$. Here, $N$
is a nucleus. Although these LFV processes take place in the MSM and DSM, none
of these processes are observed in experiments. Moreover, upper bounds are
set on the branching ratios of these processes due to non-observation of them in
experiments. In Tab. \ref{t1} we have tabulated the upper bounds obtained on the
above described LFV processes, which have been searched in experiments.
\begin{table}[!h]
\centering
\begin{tabular}{|c|c|}\hline
LFV process & current bound \\ \hline
$\mu^+\to e^+\gamma$ & $3.1\times10^{-13}$ \cite{MEGII:2023ltw} \\
$\tau^\pm\to e^\pm\gamma$ & $3.3\times10^{-8}$ \cite{BaBar:2009hkt} \\
$\tau^\pm\to \mu^\pm\gamma$ & $4.2\times10^{-8}$ \cite{Belle:2021ysv} \\
$\mu^+\to e^+e^+e^-$ & $1.0\times10^{-12}$ \cite{SINDRUM:1987nra} \\
$\tau^-\to e^-e^-e^+$ & $2.7\times10^{-8}$ \cite{Hayasaka:2010np} \\
$\tau^-\to e^-\mu^-\mu^+$ & $2.7\times10^{-8}$ \cite{Hayasaka:2010np} \\
$\tau^-\to \mu^-\mu^-e^+$ & $1.7\times10^{-8}$ \cite{Hayasaka:2010np} \\
$\tau^-\to e^-\mu^-e^+$ & $1.8\times10^{-8}$ \cite{Hayasaka:2010np} \\
$\tau^-\to e^-e^-\mu^+$ & $1.5\times10^{-8}$ \cite{Hayasaka:2010np} \\
$\tau^-\to \mu^-\mu^-\mu^+$ & $2.1\times10^{-8}$ \cite{Hayasaka:2010np} \\
$\mu^-{\rm Au}\to e^-{\rm Au}$ & $7\times10^{-13}$ \cite{SINDRUMII:2006dvw} \\
$\mu^-{\rm Ti}\to e^-{\rm Ti}$ & $4.3\times10^{-12}$ \cite{SINDRUMII:1993gxf} \\ \hline
\end{tabular}
\caption{Experimental upper bounds on various LFV processes. In the second column, the
last two entries are upper bounds on the conversion rate of $\mu\to e$ in the presence
of a nucleus. Rest of the entries in this column are upper bounds on the branching
ratios of the LFV decays.}
\label{t1}
\end{table}
The upper bounds given in Tab. \ref{t1} give constraints on the
model parameters of a model. In the present work, our aim is to study the LFV
processes and the constraints due to non-observation of these processes in the
MSM and DSM. In addition to this, we compare the contributions of the LFV processes
which give rise in these models, and that would be helpful in distinguishing
these two models
in future experiments. In our study, we have found that the 3-body muon and tau decays,
which are mentioned in Tab. \ref{t1}, can give rise to different contributions in
both these models. Hence,
there is a possibility to distinguish these two models by searching the above
mentioned decays in future experiments. On the other hand, the 2-body muon and
tau decays, which are mentioned in Tab. \ref{t1}, may give similar contributions
in both these models. The details of the above mentioned results will be described
later.

LFV processes in the MSM have already been studied previously in
\cite{Toma:2013zsa}. Below we describe the difference of our work in
comparison to that of \cite{Toma:2013zsa}. In \cite{Toma:2013zsa},
among all possible 3-body LFV decays, only the decays of the form
$\ell_\alpha\to\ell_\beta\ell_\beta\overline{\ell}_\beta$
are analyzed. After comparing with Tab. \ref{t1}, we notice
that other 3-body decays of $\ell_\tau\to\ell_\beta\ell_\delta\overline{\ell}_\delta$ and
$\ell_\tau\to\ell_\beta\ell_\beta\overline{\ell}_\delta$, where
$\beta\neq\delta$ and $\beta,\delta=e,\mu$,
are also possible in the MSM. However, these decays
are not studied in \cite{Toma:2013zsa}. Hence, in order to complete the analysis on LFV,
we have computed branching ratios of the 3-body decays
$\ell_\tau\to\ell_\beta\ell_\delta\overline{\ell}_\delta$ and
$\ell_\tau\to\ell_\beta\ell_\beta\overline{\ell}_\delta$ in the MSM and also in the DSM.
As already mentioned before, analyzing these 3-body decays is worth
to do, since these decays give
different contributions in the MSM and DSM, and hence, these models can be
distinguished by searching the LFV decays in experiments. In addition
to the above described differences, the numerical analysis of work is different
from that of \cite{Toma:2013zsa}, which we explain later. In this work,
we have analyzed all different LFV processes in the DSM for the first time. In this
context,
notice that in \cite{Guo:2020qin}, where the proposed model is closely related
to the DSM, the LFV processes
driven by the muon are studied but the 3-body tau decays are not studied.

It is described before that in the MSM and DSM, discrete symmetries $Z_2$ and
$Z_2^{(A)}\times Z_2^{(B)}$ are proposed, respectively, under which all the
standard model fields
are neutral and additional fields are charged. One of purposes of these symmetries
is to generate neutrino masses radiatively \cite{Ma:2006km,Farzan:2012sa} in
the above two models. The other purpose of these symmetries is to provide a stable
candidate in order to explain the dark matter problem \cite{Workman:2022ynf} in the
above two models. As a result of this, in our analysis on LFV in this work,
one expect us to
address the dark matter problem as well. However, it is beyond the scope of this
paper to address the problem on dark matter. Notice that in the original MSM
and the closely related model of DSM, study on the LFV and analysis on the
dark matter have been done in \cite{Kubo:2006yx} and \cite{Guo:2020qin},
respectively. Also, see \cite{Kitabayashi:2021fqx,DeRomeri:2022cem,Tapender:2024ktc}
for recent works which are based on scotogenic models in the above described direction.

The paper is organized as follows. In the next section, we describe the relevant
details of the models which are under consideration of this work. In Sec. \ref{s3}, we
describe different Feynman diagrams that give rise to LFV processes in the MSM and
DSM. In Sec. \ref{s4}, we describe our methodology of computing the amplitudes of
the LFV processes and present the analytical expressions of all the LFV processes
in the MSM and DSM.
In Sec. \ref{s6}, we present
the numerical results obtained in both the MSM and DSM. We conclude in the last
section.

\section{The models}

We have given a brief introduction to the MSM and DSM in the previous section. Here,
we describe relevant parts of these models, whose content is useful to us in
computing the LFV processes of these models.

\subsection{MSM}\label{s2.1}

Relevant fields and symmetries of the MSM \cite{Ma:2006km} are tabulated in
Tab. \ref{t2}.
\begin{table}[!h]
\centering
\begin{tabular}{|c|c|c|c|c|}\hline
Field & Spin & $SU(2)_L$ & $U(1)_Y$ & $Z_2$ \\ \hline
$L_\alpha=(\nu_\alpha,\ell_\alpha)$ & 1/2 & 2 & $-$1/2 & $+$ \\
$\ell^c_\alpha$ & 1/2 & 1 & 1 & $+$ \\
$\Phi^T=(\phi^+,\phi^0)$ & 0 & 2 & 1/2 & $+$ \\
$\eta_M^T=(\eta_M^+,\eta_M^0)$ & 0 & 2 & 1/2 & $-$ \\
$N^m_k$ & 1/2 & 1 & 0 & $-$ \\ \hline
\end{tabular}
\caption{Fields in the lepton sector of the MSM along with their charge assignments.
Here, $\Phi$ is the Higgs-like doublet.}
\label{t2}
\end{table}
Now, with these fields and symmetries, the invariant Lagrangian of the model is
\begin{equation}
-{\cal L}^M=y_{\alpha\beta}(\nu_\alpha\phi^{+^*}+\ell_\alpha\phi^{0^*})\ell^c_\beta
+f^M_{\alpha k}(\nu_\alpha\eta_M^0-\ell_\alpha\eta_M^+)N^m_k+\frac{1}{2}M_kN^m_kN^m_k+h.c.
\label{mlag}
\end{equation}
Here, $M_k$ is the mass of the Majorana field $N^M_k$, which is formed from the
Weyl fermion $N^m_k$. After the electroweak symmetry
breaking, $\Phi$ acquires vacuum expectation value (VEV), and hence, the first term
in Eq. (\ref{mlag}) gives masses to charged leptons. On the other hand, $\eta_M$ acquires
zero VEV since it is odd under the $Z_2$ symmetry, and hence, neutrinos are massless
at tree level. Neutrinos acquire masses through a radiative mechanism which is shortly
explained later.

The invariant scalar potential between $\Phi$ and $\eta_M$ is given by \cite{Ma:2006km}
\begin{eqnarray}
V_M&=&m_1^2\phi^\dagger\phi+m_2^2\eta^\dagger\eta+\frac{1}{2}\lambda_1
(\phi^\dagger\phi)^2+\frac{1}{2}\lambda_2(\eta_M^\dagger\eta_M)^2
+\lambda_3(\phi^\dagger\phi)(\eta_M^\dagger\eta_M)+\lambda_4(\phi^\dagger\eta_M)
(\eta_M^\dagger\phi)
\nonumber \\
&&+\frac{1}{2}\lambda_5[(\phi^\dagger\eta_M)^2+h.c.].
\label{mpot}
\end{eqnarray}
After minimizing this scalar potential, only the $\Phi$ should acquire VEV. Hence,
we parameterize the neutral components of $\Phi$ and $\eta_M$ as
\begin{equation}
\phi^0=\frac{H}{\sqrt{2}}+v,\quad\eta_M^0=\frac{1}{\sqrt{2}}(\eta_{MR}+i\eta_{MI})
\end{equation}
Here, $H$ is the Higgs boson and $v=$ 174 GeV. Now, after the electroweak
symmetry breaking, the mass spectrum in the MSM is given by \cite{Ma:2006km}
\begin{eqnarray}
m^2(H)\equiv m_H^2&=&2\lambda_1v^2,
\nonumber \\
m^2(\eta_M^\pm)\equiv m_{\eta_M^\pm}^2&=&m_2^2+\lambda_3v^2,
\nonumber \\
m^2(\eta_{MR})\equiv m_R^2&=&m_2^2+(\lambda_3+\lambda_4+\lambda_5)v^2,
\nonumber \\
m^2(\eta_{MI})\equiv m_I^2&=&m_2^2+(\lambda_3+\lambda_4-\lambda_5)v^2
\label{mspec}
\end{eqnarray}
The masses of the additional fields in Eq. (\ref{mspec})
have a role to play in our analysis on LFV, which we describe later in Sec. \ref{s6}.
On the other hand, using the arguments of Sec. \ref{s4}, notice that the
Higgs field $H$ does not affect the LFV processes in our work. Nevertheless, the
Higgs boson mass in the MSM can be fitted to
its experimental value with the parameter $\lambda_1$.

As stated before, neutrinos in the MSM acquire masses through a radiative
mechanism. At 1-loop level, the neutral components of $\eta_M$ and $N^M_k$ drive
masses to neutrinos in this model. By taking $\Lambda^M={\rm diag}(\Lambda^M_1,
\Lambda^M_2,\Lambda^M_3)$, the mass expressions for neutrinos at 1-loop level
can be written as \cite{Ma:2006km}
\begin{eqnarray}
({\cal M}_\nu)_{\alpha\beta}&=&(f^M\Lambda^M f^{M^T})_{\alpha\beta}=\sum_{k=1}^3
f^M_{\alpha k}f^M_{\beta k}\Lambda^M_k,
\nonumber \\
\Lambda^M_k&=&\frac{M_k}{16\pi^2}\left[\frac{m_R^2}{m_R^2-M_k^2}
\ln\frac{m_R^2}{M_k^2}-\frac{m_I^2}{m_I^2-M_k^2}\ln\frac{m_I^2}{M_k^2}\right]
\label{mnumas}
\end{eqnarray}
By diagonalizing the above mass matrix, we get the neutrino masses and also the mixing
angles. We diagonalize this matrix by parametrizing \cite{Casas:2001sr} the
Yukawa couplings as
\begin{equation}
f^M=U_{PMNS}^*\sqrt{m_\nu}R\sqrt{\Lambda^M}^{-1}
\label{mpara}
\end{equation}
Here, $U_{PMNS}$ is the Pontecorvo-Maki-Nakagawa-Sakata matrix, which can be
parameterized \cite{Workman:2022ynf} in terms of the three neutrino mixing angles,
one $CP$ violating Dirac phase and two Majorana phases.
$m_\nu$ is a diagonal matrix containing the neutrino mass eigenvalues,
which is written as $m_\nu={\rm diag}(m_1,m_2,m_3)$. $R$ is a complex
orthogonal matrix which satisfies $RR^T=I=R^TR$. Now, using the parametrization
of Eq. (\ref{mpara}), notice that
\begin{equation}
{\cal M}_\nu=U_{PMNS}^*m_\nu U_{PMNS}^\dagger
\end{equation}
From the above equation, we see that the unitary matrix that diagonalize
${\cal M}_\nu$ is $U_{PMNS}$. Hence, the mixing angles in the neutrino sector of
the MSM can be explained by parametrizing the Yukawa couplings
as given by Eq. (\ref{mpara}).

As stated in Sec. \ref{s1}, the off-diagonal elements of Yukawa couplings drive LFV
processes. Without loss of generality, the Yukawa couplings $y_{\alpha\beta}$ of
Eq. (\ref{mlag}) can be taken to be diagonal. On the other hand, the Yukawa
couplings $f^M$, whose form is given in Eq. (\ref{mpara}),
have off-diagonal elements, since $U_{PMNS}$ is off-diagonal.
As these couplings determine the strength of the LFV processes, below we give an
estimation of when these couplings become large enough. In the expression for $f^M$,
$R$ has no physical interpretation. Hence, to simplify our analysis, we take $R=I$.
As a result of this, the expression for $f^M$ becomes
\begin{equation}
f^M=U_{PMNS}^*\cdot{\rm Diag}\left(\sqrt{\frac{m_1}{\Lambda^M_1}},
\sqrt{\frac{m_2}{\Lambda^M_2}},\sqrt{\frac{m_3}{\Lambda^M_3}}\right)
\label{fm}
\end{equation}
Now we see that, the elements of $U_{PMNS}$ have magnitude of ${\cal O}(1)$. On the
other hand, the neutrino masses are tiny and hence the couplings $f^M_{\alpha k}$
can become very small. To compensate the small values in $f^M_{\alpha k}$
we can make $\Lambda^M_k$ to be very small. Notice that this is possible by taking
degenerate values for $m_R$ and $m_I$ in Eq. (\ref{mnumas}), which in turn is
possible by suppressing the coupling $\lambda_5$, and this can
be noticed from Eq. (\ref{mspec}). Finally, we have estimated that $\lambda_5$ should be
suppressed to around $10^{-10}$ in order to make $f^M_{\alpha k}\sim1$. While
making this estimation, we have taken neutrino mass scale to be around 0.1 eV
\cite{Workman:2022ynf} and the masses of additional fields to be $\sim1$ TeV.
Hence, for the above described values of parameters, LFV processes in
the MSM are unsuppressed and can have significant effects in experiments.

\subsection{DSM}\label{s2.2}

Relevant fields and symmetries of the DSM \cite{Farzan:2012sa} are tabulated in
Tab. \ref{t3}.
\begin{table}[!h]
\centering
\begin{tabular}{|c|c|c|c|c|c|c|}\hline
Field & Spin & $SU(2)_L$ & $U(1)_Y$ & $U(1)_{B-L}$ & $Z_2^{(A)}$ & $Z_2^{(B)}$ \\\hline
$L_\alpha=(\nu_\alpha,\ell_\alpha)$ & 1/2 & $2$ & $-1/2$ & $-1$ & $+$ & $+$ \\
$\ell^c_\alpha$ & 1/2 & $1$ & $1$ & $1$ & $+$ & $+$ \\
$\nu^c_\alpha$ & 1/2 & $1$ & $0$ & $1$ & $-$ & $+$ \\ \hline
$\Phi^T=(\phi^+,\phi^0)$ & 0 & $2$ & $1/2$ & $0$ & $+$ & $+$ \\
$\eta_D^T=(\eta_D^+,\eta_D^0)$ & 0 & $2$ & $1/2$ & $0$ & $+$ & $-$ \\
$\chi$ & 0 & $1$ & $0$ & $0$ & $-$ & $-$ \\ \hline
$N_k$ & 1/2 & $1$ & $0$ & $-1$ & $+$ & $-$ \\
$N_k^c$ & 1/2 & $1$ & $0$ & $1$ & $+$ & $-$ \\ \hline
\end{tabular}
\caption{Fields in the lepton sector of the DSM, along with their charge assignments.}
\label{t3}
\end{table}
Here, the symmetry $U(1)_{B-L}$ is exact \cite{Farzan:2012sa}, and without loss of
generality we have taken this to be global. The purpose of this symmetry is to have
the lepton number be conserved in the DSM. Hence, neutrinos in this model are
Dirac particles. The field $\Phi$ in Tab. \ref{t3} is the Higgs-like doublet, whereas,
$\eta_D$ and $\chi$ are the additional scalar fields of this model. Notice
that $\eta_D$ in this model is the analog of $\eta_M$ of the MSM. Moreover, after
comparing the fields between Tabs. \ref{t2} and \ref{t3}, we see that two different
Weyl fermions $N_k$ and $N_K^c$ are introduced in the DSM as against to $N^m_k$ in the
MSM. Also notice that the discrete symmetry $Z_2$ of the MSM has been modified to
$Z_2^{(A)}\times Z_2^{(B)}$ in the DSM. Now, with the charge assignments of
Tab. \ref{t3}, the allowed interaction terms in the DSM are
\begin{equation}
-{\cal L}^D=y_{\alpha\beta}(\nu_\alpha\phi^{+^*}+\ell_\alpha\phi^{0^*})\ell^c_\beta
+f^D_{k\alpha}(\nu_\alpha\eta_D^0-\ell_\alpha\eta_D^+)N^c_k
+h_{\alpha k}N_k\nu^c_\alpha\chi+M^\prime_kN_kN^c_k+h.c.
\label{dlag}
\end{equation}
Here, $M^\prime_k$ is the Dirac mass for $N^D_k=(N_k,N^c_k)$. Since $\Phi$ acquires VEV,
the first term of Eq. (\ref{dlag}) give masses to charged leptons in the DSM. On the
other hand, $\eta_D$ and $\chi$ do not acquire VEVs, since they are charged under the
$Z_2^{(B)}$, which is an exact symmetry. As a result of this, the second and third
terms of Eq. (\ref{dlag}) do not generate masses to neutrinos at tree level. Shortly later,
we explain that, due to soft breaking of the $Z_2^{(A)}$ symmetry, neutrinos acquire masses
at 1-loop level in the DSM.

The scalar potential in the DSM is \cite{Farzan:2012sa}
\begin{eqnarray}
V&=&\mu_1^{\prime^2}\Phi^\dagger\Phi+\mu_2^{\prime^2}\eta_D^\dagger\eta_D
+\frac{1}{2}\mu_3^{\prime^2}\chi^2+\frac{1}{2}
\lambda^\prime_1(\Phi^\dagger\Phi)^2+\frac{1}{2}\lambda^\prime_2(\eta_D^\dagger\eta_D)^2
+\lambda^\prime_3(\Phi^\dagger\Phi)(\eta_D^\dagger\eta_D)
\nonumber \\
&&+\lambda^\prime_4(\Phi^\dagger\eta_D)(\eta_D^\dagger\Phi)
+\frac{1}{2}[\lambda^\prime_5(\Phi^\dagger\eta_D)^2+h.c.]+\frac{1}{4}\lambda^\prime_6\chi^4
+\frac{1}{2}\lambda^\prime_7(\Phi^\dagger\Phi)\chi^2
+\frac{1}{2}\lambda^\prime_8(\eta_D^\dagger\eta_D)\chi^2
\nonumber \\
&&+A\chi[\Phi^\dagger\eta_D+h.c.]
\label{dpot}
\end{eqnarray}
Notice that we have used prime on the parameters in the above potential in order to
distinguish these with the corresponding parameters of Eq. (\ref{mpot}). In the last
term of this potential, without loss of generality, the parameter $A$ is chosen to
be real. Moreover, the parameter $\lambda_5^\prime$ is also taken to be real in order to
simplify our calculations. After minimizing
the above scalar potential, only the $\Phi$ should acquire VEV, which give rise to the
breaking of the electroweak symmetry. After this symmetry breaking, we identify the
real part of the neutral component of $\Phi$ as the Higgs boson. Moreover, neutral components
of $\eta_D$ mix with $\chi$. Hence, we write $\eta_D^0=(\eta_{DR}^0+i\eta_{DI}^0)/\sqrt{2}$.
Now, the mass spectrum of the physical scalar fields in the DSM is \cite{Farzan:2012sa}
\begin{eqnarray}
m^2(H)&\equiv &m_H^2=2\lambda^\prime_1v^2,
\nonumber \\
m^2(\eta_D^\pm)&\equiv &m_{\eta_D^\pm}^2=\mu_2^{\prime^2}+\lambda^\prime_3v^2,
\nonumber \\
m^2(\eta_{DI}^0)&\equiv &m_{\eta_{DI}^0}^2=\mu_2^{\prime^2}+(\lambda^\prime_3+\lambda^\prime_4-\lambda^\prime_5)v^2,
\nonumber \\
&&M_{\eta_{DR}^0,\chi}^2=\left(\begin{array}{cc}
m_{\eta_{DR}^0}^2 & \sqrt{2}Av \\
\sqrt{2}Av & m_{\chi}^2 \end{array}\right)
\nonumber \\
&&m_{\eta_{DR}^0}^2=\mu_2^{\prime^2}+(\lambda^\prime_3+\lambda^\prime_4+\lambda^\prime_5)v^2,
\quad m_{\chi}^2=\mu_3^{\prime^2}+\lambda^\prime_7v^2
\label{dspec}
\end{eqnarray}
Here, $M_{\eta_{DR}^0,\chi}^2$ gives the mixing masses between $\eta_{DR}^0$ and
$\chi$. After diagonalizing this, we denote the mass eigenstates as $\zeta_{1,2}$.
Moreover, the mixing between $\eta_{DR}^0$ and $\chi$ is given by
\begin{equation}
\tan2\theta=\frac{2\sqrt{2}Av}{m_{\chi}^2-m_{\eta_{DR}^0}^2}
\end{equation}
The masses of the additional fields in Eq. (\ref{dspec}) and
the mixing angle $\theta$ have a role to play in our analysis on LFV, which
we describe later in Sec. \ref{s6}. On the other hand, using the arguments of
Sec. \ref{s4},
notice that the Higgs field $H$ does not affect the LFV processes in our work. Nevertheless,
the Higgs boson mass in the DSM can be fitted to its experimental value with
the parameter $\lambda_1^\prime$.

As described before, in the DSM, neutrinos acquire masses at 1-loop level. The 1-loop
diagrams for neutrinos are driven by $N^D_k$, $\chi$ and neutral components of $\eta_D$.
In order for these loops to exist, soft breaking of the $Z_2^{(A)}$ symmetry is necessary,
which happens due to the $A$-term of Eq. (\ref{dpot}). Now, after evaluating these 1-loop
diagrams, mass expressions for neutrinos are given by \cite{Farzan:2012sa}
\begin{eqnarray}
({\cal M}_\nu)_{\alpha\beta}&=&(h\Lambda^D f^D)_{\alpha\beta}=
\sum_{k=1}^3h_{\alpha k}\Lambda^D_k f^D_{k\beta},
\nonumber \\
\Lambda^D_k&=&\frac{\sin(2\theta)}{32\pi^2\sqrt{2}}
M^\prime_k\left[\frac{m_{\zeta_1}^2}{m_{\zeta_1}^2-M^{\prime^2}_k}
\ln\frac{m_{\zeta_1}^2}{M^{\prime^2}_k}-\frac{m_{\zeta_2}^2}{m_{\zeta_2}^2-M^{\prime^2}_k}
\ln\frac{m_{\zeta_2}^2}{M^{\prime^2}_k}\right]
\label{dnumas}
\end{eqnarray}
Here, $m_{\zeta_1}$ and $m_{\zeta_2}$ are the masses of $\zeta_1$ and $\zeta_2$,
whose discussion is given previously. By diagonalizing the above mass matrix, we get
neutrino mass eigenvalues and also mixing angles. We diagonalize this matrix by
parametrizing \cite{Casas:2001sr} the Yukawa couplings as
\begin{eqnarray}
f^D&=&\sqrt{\Lambda^{D^{-1}}}S\sqrt{m_\nu}U^\dagger_{PMNS},\quad
h=V\sqrt{m_\nu}R^\dagger\sqrt{\Lambda^D}^{-1},
\nonumber \\
m_\nu&=&{\rm diag}(m_1,m_2,m_3)
\label{dpara}
\end{eqnarray}
Here, $R$ and $S$ are complex matrices which satisfy $R^\dagger S=I$. In fact, after
using Eq. (\ref{dpara}) in Eq. (\ref{dnumas}) we get $V^\dagger{\cal M}_\nu U_{PMNS}
=m_\nu$, which is the desired relation for diagonalizing the neutrino mass matrix
in the DSM. Here, $V$ is a unitary matrix which rotate the right-handed neutrino fields from
flavor to mass eigenstates. As described before, the $U_{PMNS}$ matrix is parameterized
in terms of neutrino mixing angles and $CP$ violating phases. Hence, by parametrizing
the Yukawa couplings as in Eq. (\ref{dpara}), we not only obtain the neutrino
mass eigenvalues but also the mixing angles. The above described results are analogs
of the corresponding results described in Sec. \ref{s2.1}.

Notice that, from Eq. (\ref{dpara}), the Yukawa couplings $f^D$ and $h$ can in general have
off-diagonal elements, and hence, they can drive LFV processes. However,
the couplings $h_{\alpha k}$ do not drive LFV processes of our interest,
at least at 1-loop level, since these couplings connect singlet fields.
On the other hand, the couplings $f^D_{k \alpha}$ drive LFV processes of Tab. \ref{t1}
at 1-loop level. In the expressions of $f^D$ and $h$, $R$ and $S$ have no physical
interpretations. Hence, to simplify our analysis, we take $R=S=I$. As a result of
this, the expression for $f^D$ becomes
\begin{equation}
f^D={\rm Diag}\left(\sqrt{\frac{m_1}{\Lambda^D_1}},\sqrt{\frac{m_2}{\Lambda^D_2}},
\sqrt{\frac{m_3}{\Lambda^D_3}}\right)\cdot U^\dagger_{PMNS}
\label{fd}
\end{equation}
Now, in order
not to suppress LFV processes in the DSM, we should have $f^D_{k \alpha}\sim1$. This
possibility can be achieved if $\Lambda^D_k\sim m_i$, since the elements of
$U_{PMNS}$ are of order one. As the neutrino masses $m_i$ are tiny, $\Lambda^D_k$
should be very small in order to achieve the above possibility. From the form of
$\Lambda^D_k$, which is given in
Eq. (\ref{dnumas}), we see that the above possibility becomes true if either
$\theta$ or $m_{\zeta_1}-m_{\zeta_2}$ is suppressed, for
$m_{\zeta_1},m_{\zeta_2},M^\prime_k\sim$ 1 TeV. In order to suppress
$m_{\zeta_1}-m_{\zeta_2}$, and thus
to bring degenerate masses to $\zeta_{1,2}$, we need to fine tune the parameters
of $M_{\eta_{DR}^0,\chi}^2$, which is given in Eq. (\ref{dspec}). Instead of
fine tuning several parameters of $M_{\eta_{DR}^0,\chi}^2$, we have found
it is economical to suppress $\theta$ in order to achieve the above possibility.
Since $\theta$ is the mixing angle between $\eta_{DR}^0$ and $\chi$,
notice that this angle is proportional to the $A$-parameter. As described
previously, the $A$-parameter breaks the $Z_2^{(A)}$ symmetry softly. Hence,
it is technically natural to take this parameter to be small. Finally, in order
to achieve $f^D_{k \alpha}\sim1$, $m_i\sim$ 0.1 eV and all additional fields
to have around 1 TeV masses, we have found $\theta\sim\frac{A}{v}\sim10^{-10}$.
For these values of parameters, LFV processes in the DSM can have significant
effects in experiments.

\section{LFV processes in the MSM and DSM}
\label{s3}

\begin{figure}[h]
\begin{center}

\includegraphics[]{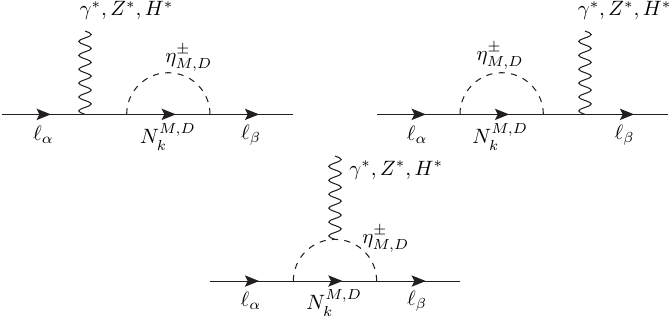}

\end{center}
\caption{Feynman diagrams of the sub-processes which lead to the decay
$\ell_\alpha\to\ell_\beta\ell_\delta\overline{\ell}_\delta$ and also the process
$\mu N\to eN$. In these diagrams, the
wavy line corresponds to off-shell photon, $Z$ or Higgs boson. On the other hand,
if the wavy line corresponds to on-shell photon, these diagrams lead to the
decay $\ell_\alpha\to\ell_\beta\gamma$. The loops in
these diagrams are driven by $N^M_k$ and $\eta^\pm_M$($N^D_k$ and $\eta^\pm_D$)
in the MSM(DSM).}
\label{f1}
\end{figure}

In this section, we describe the Feynman diagrams and various contributions to the
LFV processes in both the MSM and DSM. As discussed in Sec. \ref{s1}, the LFV processes
in the charged lepton sector of these two
models are categorized into: $\ell_\alpha\to\ell_\beta\gamma$,
$\ell_\alpha\to\ell_\beta\ell_\rho\overline{\ell}_\delta$ and $\mu N\to eN$.
All these processes take place at 1-loop level in both the MSM and DSM. Moreover,
they are driven by common sub-processes, whose Feynman diagrams are given in
Fig. \ref{f1}.
The wavy line in this figure represents an off-shell particle, which corresponds
to either photon, $Z$ or Higgs boson.
Now, by connecting the wavy line of Fig. \ref{f1} to a lepton and its
anti-lepton will give rise to the 3-body decay
$\ell_\alpha\to\ell_\beta\ell_\delta\overline{\ell}_\delta$.
Moreover, by taking $\alpha=\mu$ and $\beta=e$ and
connecting the wavy line of Fig. \ref{f1}
to quarks would lead to $\mu\to e$ conversion in the presence of a nucleus $N$.
As a result of
the above given description, diagrams of Fig. \ref{f1} can be treated as
sub-processes for the decay $\ell_\alpha\to\ell_\beta\ell_\delta\overline{\ell}_\delta$
and also for the $\mu N\to eN$ process. Notice that the
diagrams of this figure give $\gamma$-, $Z$- and Higgs-penguin contributions
to the decay $\ell_\alpha\to\ell_\beta\ell_\delta\overline{\ell}_\delta$ and
also to the process $\mu N\to eN$. In the penguin contributions, the wavy
line of Fig. \ref{f1} should represent an off-shell particle. On the other
hand, if the wavy of line of Fig. \ref{f1} represents on-shell photon,
diagrams of this figure generate the decay $\ell_\alpha\to\ell_\beta\gamma$.
Now, as indicated in Fig. \ref{f1}, the loops in this
figure are driven by either the fields of MSM or DSM, depending on the model
under consideration. As a result of this, notice
that the amplitudes of the above described LFV processes in the MSM
are related to the corresponding amplitudes of the DSM, by
interchanging the loop propagators and vertex couplings in the
amplitudes. Hence, if the couplings and the masses of propagating fields have
similar values in the MSM and DSM, the contributions to LFV processes are same
in these two models.

\begin{figure}[h]
\begin{center}

\includegraphics[]{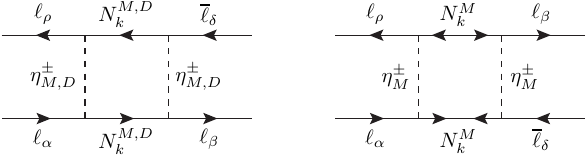}

\end{center}
\caption{Box diagrams which lead to the decay
$\ell_\alpha\to\ell_\beta\ell_\rho\overline{\ell}_\delta$. The left-hand side
diagram is possible in both the MSM and DSM, whereas, the right-hand side diagram
is possible only in the MSM. Interchanging $\ell_\beta$ with $\ell_\rho$ in these
diagrams give additional diagrams for the above decay.}
\label{f2}
\end{figure}

Apart from the penguin contributions, 3-body decays which are of the form
$\ell_\alpha\to\ell_\beta\ell_\rho\overline{\ell}_\delta$
get additional contribution due to box diagrams. These are given in
Fig. \ref{f2}.
Notice that there is no clash of arrows on the fermion propagator in the loop of the
left-hand side diagram of this figure. Hence, this diagram is possible in both the MSM and
DSM. Also notice that, the amplitude of this diagram is the same in the MSM and
DSM by interchanging the loop propagators and vertex couplings. In contrast to
the above diagram, the right-hand side diagram of Fig. \ref{f1} is possible only in the MSM.
Hence, we see that, the box diagrams give different contributions to the decay
$\ell_\alpha\to\ell_\beta\ell_\rho\overline{\ell}_\delta$ in the MSM and DSM.
Essentially, the 3-body decays in the MSM get additional contribution in
comparison to that of DSM. This result is in contrast to the penguin
contributions to the LFV processes, which is discussed in the previous
paragraph. Hence, it is likely that the MSM and DSM are distinguished
through LFV processes, by searching the decays of the form
$\ell_\alpha\to\ell_\beta\ell_\rho\overline{\ell}_\delta$ in experiments.

In general, in the charged lepton sector, a 3-body decay is of the form
$\ell_\alpha\to\ell_\beta\ell_\rho\overline{\ell}_\delta$.
Decays of this form can be categorized into:
(i) $\ell_\alpha\to\ell_\beta\ell_\beta\overline{\ell}_\beta$,
(ii) $\ell_\tau\to\ell_\beta\ell_\delta\overline{\ell}_\delta$,
(iii) $\ell_\tau\to\ell_\beta\ell_\beta\overline{\ell}_\delta$. Here, $\alpha$ can
be either $\mu$ or $\tau$ and $\alpha\neq\beta\neq\delta$. $\beta,\delta$ can take either
$e$ or $\mu$. Decays which are the forms
$\ell_\alpha\to\ell_\beta\ell_\beta\overline{\ell}_\beta$ and
$\ell_\tau\to\ell_\beta\ell_\delta\overline{\ell}_\delta$, are driven by
diagrams of Figs. \ref{f1} and \ref{f2}. On the other hand, decays
of the form $\ell_\tau\to\ell_\beta\ell_\beta\overline{\ell}_\delta$ take place
only through the diagrams of Fig. \ref{f2}. The above statements are true in
both the MSM and DSM.

\section{Analytical expressions in the MSM and DSM}
\label{s4}

In this section, we present analytical expressions for LFV
processes in the MSM and DSM. For the case of MSM, these analytical expressions are
described in Secs. \ref{s4.1} $-$ \ref{s4.5}. While for the case of DSM, we have
explained how to obtain the corresponding expressions in Sec. \ref{s4.6}.
Before we proceed further, below we describe the methodology
in obtaining the analytical expressions. As stated before,
the 3-body decays $\ell_\alpha\to\ell_\beta\ell_\rho\overline{\ell}_\delta$
may distinguish the MSM and DSM. Nevertheless, we compute the branching ratio for
$\ell_\alpha\to\ell_\beta\gamma$ and also the conversion rate for $\mu N\to eN$.
This is because, experimental limits on the processes $\mu\to e\gamma$ and
$\mu N\to eN$ are stringent and
they need to be included in the analysis of the LFV processes in the above
two models. Now, while computing the analytical expressions for the above
mentioned LFV processes, we first compute the amplitude of the decay
$\ell_\alpha\to\ell_\beta\ell_\rho\overline{\ell}_\delta$. While doing this
computation, we evaluate the penguin contribution, which is due to the diagrams of
Fig. \ref{f1}. Now, using the penguin contribution
we compute the amplitudes for $\ell_\alpha\to\ell_\beta\gamma$
and $\mu N\to eN$ processes.

\subsection{Branching ratio of $\ell_\alpha\to\ell_\beta\ell_\beta\overline{\ell}_\beta$}
\label{s4.1}

As described previously, the decay
$\ell_\alpha\to\ell_\beta\ell_\beta\overline{\ell}_\beta$
gets contribution due to diagrams of Figs.
\ref{f1} and \ref{f2}. Below we describe the penguin contribution of
Fig. \ref{f1} to the above decay and later we present the box contribution
of Fig. \ref{f2} to this decay. Now, the amplitude due to $\gamma$-penguin diagrams
to the decay
$\ell_\alpha(p_1)\to\ell_\beta(p_2)\ell_\beta(p_3)\overline{\ell}_\beta(p_4)$ is
\begin{equation}
iM_\gamma(\ell_\alpha\to\ell_\beta\ell_\beta\overline{\ell}_\beta)=
-\frac{e}{(p_1-p_2)^2}\bar{u}(p_3)\gamma_\mu v(p_4)
iM^\mu_{\ell_\alpha\to\ell_\beta\gamma^*}-(p_2\leftrightarrow p_3)
\label{mgam-a3b}
\end{equation}
Here, $M^\mu_{\ell_\alpha\to\ell_\beta\gamma^*}$ denotes the amplitude of
the sub-process $\ell_\alpha\to\ell_\beta\gamma^*$. This amplitude is
the resultant of the sum of the contributions of the diagrams of Fig. \ref{f1}.
Each of these diagrams give a divergent quantity. However, these divergences
cancel out after summing the contributions of these diagrams. In the next
paragraph, we describe the method \cite{Grimus:2002ux} we have used for getting
a finite amplitude from the diagrams of Fig. \ref{f1}.

We assign momenta $p_1$,
$p$ and $p_2$ to $\ell_\alpha$, $N^M_k$ and $\ell_\beta$,
respectively, in the diagrams of Fig. \ref{f1}. We define the following
quantities, which appear in the propagators of the loops of Fig. \ref{f1}.
\begin{equation}
D_k=p^2-M_k^2,\quad D_{1\eta}=(p-p_1)^2-m_{\eta^\pm_M}^2,\quad
D_{2\eta}=(p-p_2)^2-m_{\eta^\pm_M}^2
\end{equation}
Now, we define $b_{1,2}^k$, $c_{1,2}^k$, $d_{1,2}^k$, $f^k$ and $u^k$ through
the following integrals \cite{Grimus:2002ux}:
\begin{eqnarray}
&&\int\frac{d^dp}{(2\pi)^d}\frac{p^\mu}{D_kD_{1\eta}}=b_1^kp_1^\mu,\quad
\int\frac{d^dp}{(2\pi)^d}\frac{p^\mu}{D_kD_{2\eta}}=b_2^kp_2^\mu,
\nonumber \\
&&\int\frac{d^dp}{(2\pi)^d}\frac{p^\mu}{D_kD_{1\eta}D_{2\eta}}=
c_1^kp_1^\mu+c_2^kp_2^\mu,
\nonumber \\
&&\int\frac{d^dp}{(2\pi)^d}\frac{p^\mu p^\nu}{D_kD_{1\eta}D_{2\eta}}=
d_1^kp_1^\mu p_1^\nu+d_2^kp_2^\mu p_2^\nu+f^k(p_1^\mu p_2^\nu+p_2^\mu p_1^\nu)
+u^kg^{\mu\nu}
\label{int}
\end{eqnarray}
Notice that the above integrals appear in the amplitudes of the diagrams of Fig.
\ref{f1}. Here, the dimension $d$ is taken to be 4, after the end of loop
calculations. Using dimensional arguments, we see that $b_{1,2}^k$ and $u^k$ are
divergent quantities, whereas , $c_{1,2}^k$, $d_{1,2}^k$ and $f^k$
are finite. Now, using the integrals of Eq. (\ref{int}), we obtain the following
relations \cite{Grimus:2002ux}:
\begin{eqnarray}
&&b_1^k-b_2^k=q^2(d_1^k-d_2^k)+
(m_{\ell_\alpha}^2-m_{\ell_\beta}^2)(\kappa_1^k+\kappa_2^k),
\label{b1-b2}
\\
&&m_{\ell_\alpha}^2b_1^k-m_{\ell_\beta}^2b_2^k=
q^2(m_{\ell_\alpha}^2d_1^k-m_{\ell_\beta}^2d_2^k)
+(m_{\ell_\alpha}^2-m_{\ell_\beta}^2)[2u^k-q^2f^k+
m_{\ell_\alpha}^2\kappa_1^k+m_{\ell_\beta}^2\kappa_2^k],
\nonumber \\
\label{mb1-mb2}
\\
&&\kappa_1^k=d_1^k+f^k-c_1^k,\quad\kappa_2^k=d_2^k+f^k-c_2^k
\end{eqnarray}
Here, $q=p_1-p_2$.
As stated before, in the integrals of Eq. (\ref{int}), $b_{1,2}^k$ and $u^k$
are divergent quantities and $c_{1,2}^k$, $d_{1,2}^k$ and $f^k$ are finite. Since
these integrals appear in the amplitudes of the diagrams of Fig. \ref{f1},
after using the relations Eqs. (\ref{b1-b2}) and (\ref{mb1-mb2}) and
after summing the contributions of these diagrams, we notice
that divergences cancel out. As a result
of this, after summing the contributions of these diagrams, the amplitude
depends on the finite quantities $c_{1,2}^k$, $d_{1,2}^k$ and $f^k$. Below
we give expressions for these quantities, which can be found by using Eq.
(\ref{int}). While obtaining these expressions, we neglect the masses of
charged leptons. Hence, we have found
\begin{eqnarray}
c^k=c_1^k=c_2^k&=&\frac{-i}{16\pi^2}\int_0^1dx\int_0^{1-x}dy
\frac{y}{xM_K^2+(1-x)m^2_{\eta^\pm_M}}
\nonumber \\
d^k=d_1^k=d_2^k&=&\frac{-i}{16\pi^2}\int_0^1dx\int_0^{1-x}dy
\frac{y^2}{xM_K^2+(1-x)m^2_{\eta^\pm_M}}
\nonumber \\
f^k&=&\frac{-i}{16\pi^2}\int_0^1dx\int_0^{1-x}dy
\frac{y(1-x-y)}{xM_K^2+(1-x)m^2_{\eta^\pm_M}}
\end{eqnarray}
Since $c_1^k=c_2^k$ and $d_1^k=d_2^k$, we get $\kappa_1^k=\kappa_2^k=\kappa^k$.
The description given above in this paragraph is applicable irrespective of
the wavy line of Fig. \ref{f1} corresponds to $\gamma^*$, $Z^*$ or $H^*$.

Let the wavy line of Fig. \ref{f1} be off-shell $\gamma$. Now, after using
the description of the previous paragraph, we have found the
amplitude for the sub-process $\ell_\alpha\to\ell_\beta\gamma^*$ as
\begin{eqnarray}
iM_{\ell_\alpha\to\ell_\beta\gamma^*}^\mu&=&ie\bar{u}(p_2)
[A_1^{(\alpha,\beta)}q^2\gamma^\mu P_L
+A_2^{(\alpha,\beta)}(m_{\ell_\alpha}P_R+m_{\ell_\beta}P_L)i\sigma^{\mu\nu}q_\nu
\nonumber \\
&&-A_1^{(\alpha,\beta)}(m_{\ell_\alpha}P_R-m_{\ell_\beta}P_L)q^\mu]u(p_1),
\nonumber \\
A_1^{(\alpha,\beta)}&=&\sum_{k=1}^3f^M_{\alpha k}f^{M^*}_{\beta k}(-i)(d^k-f^k)
=-\sum_{k=1}^3f^M_{\alpha k}f^{M^*}_{\beta k}\frac{1}{16\pi^2}
\frac{1}{6m^2_{\eta^\pm_M}}G_1(x_k),
\nonumber \\
A_2^{(\alpha,\beta)}&=&-\sum_{k=1}^3f^M_{\alpha k}f^{M^*}_{\beta k}(-i)\kappa^k
=-\sum_{k=1}^3f^M_{\alpha k}f^{M^*}_{\beta k}\frac{1}{16\pi^2}
\frac{1}{2m^2_{\eta^\pm_M}}G_2(x_k),
\nonumber \\
x_k&=&\frac{M_k^2}{m^2_{\eta^\pm_M}},
\nonumber \\
G_1(x)&=&\frac{1}{6(1-x)^4}[2-9x+18x^2-11x^3+6x^3\ln x],
\nonumber \\
G_2(x)&=&\frac{1}{6(1-x)^4}[1-6x+3x^2+2x^3-6x^2\ln x]
\label{mmugam}
\end{eqnarray}
Here, $q=p_1-p_2$, which is the momentum transfer in $\gamma^*$.
$m_{\ell_\alpha}$ is the mass of charged lepton $\ell_\alpha$.
$P_{L,R}=\frac{1\mp\gamma_5}{2}$. Notice that,
$M_{\ell_\alpha\to\ell_\beta\gamma^*}^\mu$ satisfies the Ward identity
$q_\mu M_{\ell_\alpha\to\ell_\beta\gamma^*}^\mu=0$.
After substituting the above amplitude
in Eq. (\ref{mgam-a3b}), we get the amplitude for the decay
$\ell_\alpha\to\ell_\beta\ell_\beta\overline{\ell}_\beta$ due to the
$\gamma$-penguin diagrams.

Above, we have described the calculations which lead to the amplitude for
$\ell_\alpha\to\ell_\beta\ell_\beta\overline{\ell}_\beta$ due to the
$\gamma$-penguin diagrams. In analogy to this, we compute the amplitude
for the above decay due to the $Z$-penguin diagrams. This amplitude is
given below, after neglecting the charged lepton masses.
\begin{equation}
iM_Z(\ell_\alpha\to\ell_\beta\ell_\beta\overline{\ell}_\beta)=
\frac{g}{\cos\theta_Wm_Z^2}\bar{u}(p_3)\gamma_\mu
(\frac{1}{2}P_L-\sin^2\theta_W)v(p_4)iM^\mu_{\ell_\alpha\to\ell_\beta Z^*}
-(p_2\leftrightarrow p_3)
\label{mz-a3b}
\end{equation}
Here, $g$ is the $SU(2)_L$ gauge coupling, $\theta_W$ is the weak mixing angle
and $m_Z$ is the mass of $Z$ gauge boson. $M^\mu_{\ell_\alpha\to\ell_\beta Z^*}$
denotes the amplitude for the sub-process $\ell_\alpha(p_1)\to\ell_\beta(p_2)Z^*(q)$,
whose form is given below.
\begin{eqnarray}
iM^\mu_{\ell_\alpha\to\ell_\beta Z^*}&=&i\frac{g}{\cos\theta_W}
\bar{u}(p_2)\left[(\frac{1}{2}-\sin^2\theta_W)A_1^{(\alpha,\beta)}q^2\gamma^\mu P_L
+A_2^{(\alpha,\beta)}m_{\ell_\alpha}m_{\ell_\beta}\gamma^\mu P_R
\right.\nonumber \\
&&\left.
+(\frac{1}{2}-\sin^2\theta_W)A_2^{(\alpha,\beta)}(m_{\ell_\alpha}P_R+m_{\ell_\beta}P_L)
i\sigma^{\mu\nu}q_\nu
\right.\nonumber \\
&&\left.
-(\frac{1}{2}-\sin^2\theta_W)A_1^{(\alpha,\beta)}(m_{\ell_\alpha}P_R-m_{\ell_\beta}P_L)
q^\mu\right]u(p_1)
\label{mmuz}
\end{eqnarray}
See that the above amplitude and that in Eq. (\ref{mmugam}) have similar forms.
Because of this similarity, the amplitude in Eq. (\ref{mz-a3b}) is suppressed
in comparison to that of Eq. (\ref{mgam-a3b}), due to $Z$ boson propagator mass.
As a result of this, we have neglected the contribution due to $Z$-penguin
diagrams in the computation of branching ratio of
$\ell_\alpha\to\ell_\beta\ell_\beta\overline{\ell}_\beta$.

In analogy to the above given description, we can compute the amplitude to
$\ell_\alpha\to\ell_\beta\ell_\beta\overline{\ell}_\beta$ due to Higgs-penguin
diagrams, by taking the wavy line of Fig. \ref{f1} as off-shell Higgs boson. We expect
this amplitude be sub-dominant due to the following reasons. First,
in the Higgs-penguin diagrams, the Higgs boson always connects to either
$\ell_e\overline{\ell}_e$ or $\ell_\mu\overline{\ell}_\mu$, whose coupling
strength is very small in comparison to that of gauge coupling. Second,
in the amplitude, we expect to get propagator suppression due to Higgs boson
mass, which is in analogy to what we have found for $Z$-penguin diagrams.
Because of the above two reasons, we neglect the contribution due to
Higgs-penguin diagrams to the decay
$\ell_\alpha\to\ell_\beta\ell_\beta\overline{\ell}_\beta$.

After computing the penguin contributions, we now describe the
box contributions to $\ell_\alpha\to\ell_\beta\ell_\beta\overline{\ell}_\beta$.
As stated previously, diagrams of Fig. \ref{f2} give box contributions,
which are finite. The amplitude due to these contributions to the above decay
is
\begin{eqnarray}
iM_{\rm box}(\ell_\alpha\to\ell_\beta\ell_\beta\overline{\ell}_\beta)&=&
ie^2B^{(\beta,\beta,\overline{\beta})}_\alpha
\bar{u}(p_3)\gamma^\mu P_Lv(p_4)\bar{u}(p_2)\gamma_\mu P_Lu(p_1)
\nonumber \\
e^2B^{(\beta,\beta,\overline{\beta})}_\alpha&=&
B_{1\alpha}^{(\beta,\beta,\overline{\beta})}+B_{2\alpha}^{(\beta,\beta,\overline{\beta})},
\nonumber \\
B_{1\alpha}^{(\beta,\beta,\overline{\beta})}
&=&-\frac{1}{16\pi^2m^2_{\eta^\pm_M}}\sum_{n,k=1}^3
f^M_{\alpha n}f^{M^*}_{\beta n}f^M_{\beta k}f^{M^*}_{\beta k}D_1(x_n,x_k)
\nonumber \\
B_{2\alpha}^{(\beta,\beta,\overline{\beta})}
&=&\frac{1}{16\pi^2m^2_{\eta^\pm_M}}\sum_{n,k=1}^3
f^M_{\alpha n}f^{M}_{\beta n}f^{M^*}_{\beta k}f^{M^*}_{\beta k}\sqrt{x_nx_k}D_2(x_n,x_k)
\nonumber \\
D_1(x,y)&=&\frac{1}{2(1-x)(1-y)}+\frac{x^2\ln x}{2(1-x)^2(x-y)}
+\frac{y^2\ln y}{2(1-y)^2(y-x)}
\nonumber \\
D_2(x,y)&=&-\frac{1}{(1-x)(1-y)}-\frac{x\ln x}{(1-x)^2(x-y)}
-\frac{y\ln y}{(1-y)^2(y-x)}
\label{mbox}
\end{eqnarray}
Here, $B_{1\alpha}^{(\beta,\beta,\overline{\beta})}$ and
$B_{2\alpha}^{(\beta,\beta,\overline{\beta})}$ are the contributions arising
due to left- and right-hand side diagrams of Fig. \ref{f2}, respectively.

We have described the amplitudes arising due to penguin and box diagrams to the
decay $\ell_\alpha\to\ell_\beta\ell_\beta\overline{\ell}_\beta$. As stated
above, we are neglecting contributions due to $Z$- and Higgs-penguin
diagrams. Now, after squaring the above described amplitudes, we get the
branching ratio for the decay $\ell_\alpha\to\ell_\beta\ell_\beta\overline{\ell}_\beta$,
which is
\begin{eqnarray}
{\rm Br}(\ell_\alpha\to\ell_\beta\ell_\beta\overline{\ell}_\beta)&=&
\frac{3(4\pi\alpha_{em})^2}{8G_F^2}\left\{\frac{1}{6}
|B^{(\beta,\beta,\overline{\beta})}_\alpha|^2
+|A_1^{(\alpha,\beta)}|^2+|A_2^{(\alpha,\beta)}|^2(\frac{16}{3}\ln\frac
{m_{\ell_\alpha}}{2m_{\ell_\beta}}-\frac{26}{9})
\right. \nonumber \\
&&\left. +[-\frac{1}{3}B^{(\beta,\beta,\overline{\beta})}_\alpha(A_1^{(\alpha,\beta)})^*
+\frac{2}{3}B^{(\beta,\beta,\overline{\beta})}_\alpha(A_2^{(\alpha,\beta)})^*
-2A_1^{(\alpha,\beta)}(A_2^{(\alpha,\beta)})^*+h.c.]\right\}
\nonumber \\
&&\times{\rm Br}(\ell_\alpha\to\ell_\beta\nu_\alpha\overline{\nu}_\beta)
\label{bra3b}
\end{eqnarray}
Here, $\alpha_{em}=\frac{e^2}{4\pi}$ and $G_F$ is the Fermi constant.

In Sec. \ref{s1}, we have stated that, in the MSM, study on the LFV processes
have previously been done in \cite{Toma:2013zsa}. Here we compare the
analytical results we have obtained for
$\ell_\alpha\to\ell_\beta\ell_\beta\overline{\ell}_\beta$ in comparison
to that of \cite{Toma:2013zsa}. We notice that the expressions
for the amplitudes we have given in Eqs. (\ref{mgam-a3b}) and (\ref{mbox})
agree with the corresponding expressions of \cite{Toma:2013zsa}. In fact,
the dipole and non-dipole contributions, which are given in \cite{Toma:2013zsa},
have an overall minus sign in comparison to that of ours. But otherwise,
the form of these contributions agree with that of ours. However,
we have found some discrepancy in the amplitude of Eq. (\ref{mz-a3b})
with the corresponding expression of \cite{Toma:2013zsa}. Nevertheless, we
have argued that $Z$-penguin diagrams give sub-dominant contribution to
the decay $\ell_\alpha\to\ell_\beta\ell_\beta\overline{\ell}_\beta$, which
is also the case in \cite{Toma:2013zsa}. As a result of this, we expect
that, our analytical expressions for
$\ell_\alpha\to\ell_\beta\ell_\beta\overline{\ell}_\beta$
agree with that of \cite{Toma:2013zsa}. Indeed, the branching ratio
of this decay, which is given in Eq. (\ref{bra3b}), agrees with
the corresponding expression of \cite{Toma:2013zsa}, apart from a minor difference
in the term multiplying $|A_2^{(\alpha,\beta)}|^2$.

In the numerical analysis, which we present in Sec. \ref{s6}, we denote the
decay $\ell_\alpha\to\ell_\beta\ell_\beta\overline{\ell}_\beta$ as
$\ell_\alpha\to3\ell_\beta$, for the sake of simplicity.

\subsection{Branching ratio of
$\ell_\tau\to\ell_\beta\ell_\delta\overline{\ell}_\delta$}
\label{s4.2}

The decay
$\ell_\tau\to\ell_\beta\ell_\delta\overline{\ell}_\delta$
gets contribution due to both penguin and box diagrams. Computation of
these contributions to the above decay
is analogous to what we have presented in the previous subsection.
Now, the amplitude to the decay
$\ell_\tau(p_1)\to\ell_\beta(p_2)\ell_\delta(p_3)\overline{\ell}_\delta(p_4)$
due to $\gamma$-penguin diagrams is
\begin{equation}
iM_\gamma(\ell_\tau\to\ell_\beta\ell_\delta\overline{\ell}_\delta)=
-\frac{e}{(p_1-p_2)^2}\bar{u}(p_3)\gamma_\mu v(p_4)
iM^\mu_{\ell_\tau\to\ell_\beta\gamma^*}
\label{mgam-t}
\end{equation}
Here, $M^\mu_{\ell_\tau\to\ell_\beta\gamma^*}$ is the amplitude for the
sub-process $\ell_\tau\to\ell_\beta\gamma^*$, which can be found from
$M^\mu_{\ell_\alpha\to\ell_\beta\gamma^*}$ by putting $\alpha=\tau$
in Eq. (\ref{mmugam}). Now, the amplitude to the above decay due to
$Z$-penguin contribution can be analogously found from Eqs. (\ref{mz-a3b})
and (\ref{mmuz}). However, notice that this contribution is sub-dominant
due to propagator suppression of $Z$ boson mass. Hence, this contribution is
neglected in the rest of the calculations. Similarly, the Higgs-penguin
contribution is also neglected for the above decay, due to the reasons
presented in the previous subsection. Now, the amplitude due to box diagrams
of Fig. \ref{f2} to the above decay is given by
\begin{eqnarray}
iM_{\rm box}(\ell_\tau\to\ell_\beta\ell_\delta\overline{\ell}_\delta)&=&
ie^2B^{(\beta,\delta,\overline{\delta})}_\tau
\bar{u}(p_3)\gamma^\mu P_Lv(p_4)\bar{u}(p_2)\gamma_\mu P_Lu(p_1)
\nonumber \\
e^2B^{(\beta,\delta,\overline{\delta})}_\tau&=&
B_{1\tau}^{(\beta,\delta,\overline{\delta})}+B_{2\tau}^{(\beta,\delta,\overline{\delta})},
\nonumber \\
B_{1\tau}^{(\beta,\delta,\overline{\delta})}
&=&-\frac{1}{32\pi^2m^2_{\eta^\pm_M}}\sum_{n,k=1}^3
[f^M_{\tau n}f^{M^*}_{\beta n}f^M_{\delta k}f^{M^*}_{\delta k}
+f^M_{\delta n}f^{M^*}_{\beta n}f^M_{\tau k}f^{M^*}_{\delta k}]D_1(x_n,x_k)
\nonumber \\
B_{2\tau}^{(\beta,\delta,\overline{\delta})}
&=&\frac{1}{16\pi^2m^2_{\eta^\pm_M}}\sum_{n,k=1}^3
f^M_{\tau n}f^{M}_{\delta n}f^{M^*}_{\delta k}f^{M^*}_{\beta k}\sqrt{x_nx_k}D_2(x_n,x_k)
\label{mbox-t}
\end{eqnarray}
Here again, $B_1^{(\beta,\delta)}$ and $B_2^{(\beta,\delta)}$ are the contributions arising
due to left- and right-hand side diagrams of Fig. \ref{f2}, respectively.

After squaring the amplitudes in Eqs. (\ref{mgam-t}) and (\ref{mbox-t}), we
get branching ratio of the decay
$\ell_\tau\to\ell_\beta\ell_\delta\overline{\ell}_\delta$, which is
\begin{eqnarray}
{\rm Br}(\ell_\tau\to\ell_\beta\ell_\delta\overline{\ell}_\delta)&=&
\frac{3(4\pi\alpha_{em})^2}{8G_F^2}\left\{\frac{1}{3}
|B^{(\beta,\delta,\overline{\delta})}_\tau|^2
+\frac{2}{3}|A_1^{(\tau,\beta)}|^2+|A_2^{(\tau,\beta)}|^2(\frac{16}{3}\ln\frac
{m_{\ell_\tau}}{2m_{\ell_\delta}}-\frac{32}{9})
\right. \nonumber \\
&&\left. +[-\frac{1}{3}B^{(\beta,\delta,\overline{\delta})}_\tau(A_1^{(\tau,\beta)})^*
+\frac{2}{3}B^{(\beta,\delta,\overline{\delta})}_\tau(A_2^{(\tau,\beta)})^*
-\frac{4}{3}A_1^{(\tau,\beta)}(A_2^{(\tau,\beta)})^*+h.c.]\right\}
\nonumber \\
&&\times{\rm Br}(\ell_\tau\to\ell_\beta\nu_\tau\overline{\nu}_\beta)
\label{brtd}
\end{eqnarray}

\subsection{Branching ratio of
$\ell_\tau\to\ell_\beta\ell_\beta\overline{\ell}_\delta$}
\label{s4.3}

The decay
$\ell_\tau(p_1)\to\ell_\beta(p_2)\ell_\beta(p_3)\overline{\ell}_\delta(p_4)$
takes place only through the box diagrams of Fig. \ref{f2}.
The amplitude for this decay is
\begin{eqnarray}
iM_{\rm box}(\ell_\tau\to\ell_\beta\ell_\beta\overline{\ell}_\delta)&=&
ie^2B^{(\beta,\beta,\overline{\delta})}_\tau
\bar{u}(p_3)\gamma^\mu P_Lv(p_4)\bar{u}(p_2)\gamma_\mu P_Lu(p_1)
\nonumber \\
e^2B^{(\beta,\beta,\overline{\delta})}_\tau&=&
B_{1\tau}^{(\beta,\beta,\overline{\delta})}+B_{2\tau}^{(\beta,\beta,\overline{\delta})},
\nonumber \\
B_{1\tau}^{(\beta,\beta,\overline{\delta})}
&=&-\frac{1}{16\pi^2m^2_{\eta^\pm_M}}\sum_{n,k=1}^3
f^M_{\tau n}f^{M^*}_{\beta n}f^M_{\delta k}f^{M^*}_{\beta k}D_1(x_n,x_k)
\nonumber \\
B_{2\tau}^{(\beta,\beta,\overline{\delta})}
&=&\frac{1}{16\pi^2m^2_{\eta^\pm_M}}\sum_{n,k=1}^3
f^M_{\tau n}f^{M}_{\delta n}f^{M^*}_{\beta k}f^{M^*}_{\beta k}\sqrt{x_nx_k}D_2(x_n,x_k)
\end{eqnarray}
After squaring this amplitude, the branching ratio of the above decay is
\begin{equation}
{\rm Br}(\ell_\tau\to\ell_\beta\ell_\beta\overline{\ell}_\delta)=
\frac{3(4\pi\alpha_{em})^2}{8G_F^2}\left\{\frac{1}{6}
|B^{(\beta,\beta,\overline{\delta})}_\tau|^2\right\}
\times{\rm Br}(\ell_\tau\to\ell_\beta\nu_\tau\overline{\nu}_\beta)
\label{brtb}
\end{equation}

\subsection{Branching ratio of $\ell_\alpha\to\ell_\beta\gamma$}
\label{s4.4}

Previously, we have described the amplitude for the process
$\ell_\alpha\to\ell_\beta\gamma^*$ in Eq. (\ref{mmugam}). By taking $\gamma$
to be on-shell in this process, one can read out the amplitude for
$\ell_\alpha(p_1)\to\ell_\beta(p_2)\gamma(q)$, which is given below.
\begin{equation}
iM(\ell_\alpha\to\ell_\beta\gamma)=ie\bar{u}(p_2)
A_2^{(\alpha,\beta)}(m_{\ell_\alpha}P_R+m_{\ell_\beta}P_L)i\sigma^{\mu\nu}q_\nu
\epsilon_\mu(q) u(p_1)
\end{equation}
After squaring this amplitude, we get the branching ratio for the above decay,
which is
\begin{equation}
{\rm Br}(\ell_\alpha\to\ell_\beta\gamma)=\frac{48\pi^3\alpha_{em}}{G_F^2}
|A_2^{(\alpha,\beta)}|^2{\rm Br}(\ell_\alpha\to\ell_\beta\nu_\alpha\overline{\nu}_\beta)
\end{equation}
The above branching ratio formula agrees with the corresponding formula,
which is originally given in \cite{Kubo:2006yx}.

\subsection{Conversion rate of $\mu\to e$ in a nucleus}
\label{s4.5}

As described in Sec. \ref{s3}, $\mu\to e$ conversion in a nucleus takes place
through the penguin diagrams of Fig. \ref{f1}, by attaching the wavy line of
this figure to a quark field. Now, analogous to the arguments given for the
case of 3-body decays, we expect $\gamma$-penguin diagrams to give dominant
contribution to the $\mu\to e$ conversion, since the $Z$- and Higgs-penguin
diagrams have propagator suppressions.
The theory for computing conversion rate of $\mu\to e$ in a nucleus is described
in \cite{Kuno:1999jp}. The analytical expression for this conversion rate is
\cite{Kuno:1999jp}
\begin{eqnarray}
{\rm CR}(\mu\to e,{\rm Nucleus})&=&
\frac{p_eE_em_{\ell_\mu}^3G_F^2\alpha_{em}^3Z_{eff}^4F_p^2}{8\pi^2Z}
\nonumber \\
&&\times\left\{|(Z+N)(g^{(0)}_{LV}+g^{(0)}_{LS})+(Z-N)(g^{(1)}_{LV}+g^{(1)}_{LS})|^2
\right.\nonumber \\
&&\left.+|(Z+N)(g^{(0)}_{RV}+g^{(0)}_{RS})+(Z-N)(g^{(1)}_{RV}+g^{(1)}_{RS})|^2\right\}
\frac{1}{\Gamma_{\rm capt}}
\end{eqnarray}
Here, $p_e$ and $E_e$ are the momentum and energy of the electron.
$Z_{eff}$ and $F_p$ are the effective atomic charge and nuclear matrix element,
whose definitions can be found in \cite{Kuno:1999jp}.
$\Gamma_{\rm capt}$ is the total muon capture rate. $Z$ and $N$ are the
number of protons and neutrons in the nucleus. The quantities $g^{(0)}_{XK}$
and $g^{(1)}_{XK}$, where $X=L,R$ and $K=S,V$, are isoscalar and isovector
couplings, respectively. These couplings arise due to mediation of either
scalar or vector field between quarks and the effective $\mu-e$ vertex.
Numerical values for the quantities $p_e$, $Z_{eff}$, $F_p$ and
$\Gamma_{\rm capt}$ can be found in \cite{Chiang:1993xz,Kosmas:2001mv,
Kitano:2002mt} for various nuclei.
To compute the couplings $g^{(0)}_{XK}$ and $g^{(1)}_{XK}$, we
have followed \cite{Arganda:2007jw}. Below, we briefly describe the computation
of these couplings. The effective $\mu-e$ vertex, in our work, is generated
by the diagrams of Fig. \ref{f1}. Since we have argued that Higgs-penguin
diagrams give sub-dominant contribution, we have taken
$g^{(0)}_{XS}= 0= g^{(1)}_{XS}$. We have also neglected the contribution
due to $Z$-penguin diagrams in the effective $\mu-e$ vertex. Now, the
amplitude due to
$\gamma$-penguin diagrams is given in Eq. (\ref{mmugam}), which gives a
parametrization to the effective $\gamma\mu e$ vertex. Notice that the
term containing $q^\mu$ in Eq. (\ref{mmugam}) does not contribute to the
$\mu\to e$ conversion at the quark level. Now, following the procedure
described in \cite{Arganda:2007jw}, we have found
\begin{eqnarray}
g^{(0)}_{LV}=g^{(1)}_{LV}&=&\frac{4\pi\alpha_{em}}{\sqrt{2}G_F}
(A_1^{(\mu,e)}-A_2^{(\mu,e)})
\nonumber \\
g^{(0)}_{RV}=g^{(1)}_{RV}&=&-\frac{4\pi\alpha_{em}}{\sqrt{2}G_F}
\frac{m_{\ell_e}}{m_{\ell_\mu}}A_2^{(\mu,e)}
\end{eqnarray}

\subsection{Analytical expressions in the DSM}
\label{s4.6}

In Sec. \ref{s3}, we have described the Feynman diagrams which induce
the LFV processes in the MSM and DSM. We have explained that
the amplitudes of LFV processes in the DSM can
be found from the corresponding expressions of the MSM, by replacing
the vertex couplings and propagator masses accordingly. As a result of
this, in the case of LFV processes $\ell_\alpha\to\ell_\beta\gamma$ and
$\mu\to e$ conversion, the analytical expressions presented in Secs. \ref{s4.4} and
\ref{s4.5} are valid for the DSM, by making the following replacements
in these expressions:
\begin{equation}
f^M_{\alpha k}\to f^D_{k\alpha},\quad m^2_{\eta^\pm_M}\to m^2_{\eta^\pm_D},
\quad M_k\to M_k^\prime
\label{rep}
\end{equation}
Now, for the case of 3-body LFV decays, it is explained in Sec. \ref{s3}
that the decays in the MSM get additional contribution due to right-hand
side diagram of Fig. \ref{f2}. As a result of this, while obtaining analytical
expressions for 3-body LFV decays in the DSM, one should discard
the contributions of $B_{2\alpha}^{(\beta,\beta,\overline{\beta})}$,
$B_{2\tau}^{(\beta,\delta,\overline{\delta})}$ and
$B_{2\tau}^{(\beta,\beta,\overline{\delta})}$, which are presented
in Secs. \ref{s4.1}, \ref{s4.2} and \ref{s4.3}, respectively. Now, after
doing this, the analytical expressions presented in
Secs. \ref{s4.1}, \ref{s4.2} and \ref{s4.3} are applicable for the case
of DSM, by making the replacements of Eq. (\ref{rep}).

\section{Numerical analysis}
\label{s6}

After presenting analytical expressions in the previous two sections, here we present
numerical analysis of the LFV processes in both the MSM and DSM. Among all the LFV
processes, which are listed in Tab. \ref{t1}, the most stringent experimental limits
are obtained on ${\rm Br}(\mu\to e\gamma)$, ${\rm Br}(\mu\to3e)$ and
${\rm CR}(\mu\to e,{\rm Au})$. Hence, we first do an analysis on the above three observables
and study which of these observables can give dominant constraints on the parameter
space of the MSM and DSM. While doing the above analysis, we have found that the
perturbativity
of Yukawa couplings play a major role. The above analysis is relevant in work, since
the experimental limits on the above three observables need to be satisfied while
doing the analysis on the LFV $\tau$ decays.

As mentioned previously, studies on LFV processes in the MSM and DSM have been
done in \cite{Toma:2013zsa,Guo:2020qin}. Below we describe the difference between our
analysis, which is described in the previous paragraph, and that of
\cite{Toma:2013zsa,Guo:2020qin}. In \cite{Toma:2013zsa}, it is studied under
what conditions ${\rm Br}(\mu\to3e)$ and ${\rm CR}(\mu\to e,{\rm Au})$ can exceed
over ${\rm Br}(\mu\to e\gamma)$,
by considering specific benchmark points of the parameter space of the MSM. Notice
that the experimental limits on ${\rm Br}(\mu\to e\gamma)$, ${\rm Br}(\mu\to3e)$
and ${\rm CR}(\mu\to e,{\rm Au})$
are different. Hence, it is more reasonable to study the following question: if the
experimental limit on ${\rm Br}(\mu\to e\gamma)$ is satisfied, is there any parameter
space where ${\rm Br}(\mu\to3e)$ and ${\rm CR}(\mu\to e,{\rm Au})$ can exceed over
the experimental limits on these observables? The above
question is addressed in this work by considering the role of perturbativity of
Yukawa couplings. In \cite{Toma:2013zsa}, perturbativity bounds on Yukawa couplings
are satisfied in the analysis. However, the role of these perturbativity bounds
is not discussed. On the other hand, in \cite{Guo:2020qin}, where the model is
closely related to the DSM, a correlation between ${\rm Br}(\mu\to e\gamma)$ and
${\rm Br}(\mu\to3e)$ is studied. However, our analysis is more general than that
of \cite{Guo:2020qin}.

As stated before, perturbativity bounds on Yukawa couplings play a major role in
the analysis of LFV in the MSM and DSM. In order that the Yukawa couplings to be
within the perturbative bounds, we should have
\begin{equation}
|f^M_{\alpha k}|\leq\sqrt{4\pi},\quad |f^D_{k\alpha}|\leq\sqrt{4\pi}
\label{perb}
\end{equation}
in both the MSM and DSM, respectively. The expressions for $f^M$ and $f^D$, in the
case of our analysis, have been given in Eqs. (\ref{fm}) and (\ref{fd}), respectively.
We see that the magnitudes of the elements of $U_{PMNS}$ are less than ${\cal O}(1)$.
As a result of this, we define the following quantity
\begin{equation}
{\rm PER}_{M(D)}={\rm Max}\left[\sqrt{\frac{m_1}{4\pi|\Lambda^{M(D)}_1|}},
\sqrt{\frac{m_2}{4\pi|\Lambda^{M(D)}_2|}},\sqrt{\frac{m_3}{4\pi|\Lambda^{M(D)}_3|}}\right]
\end{equation}
We now see that, if PER$_M\leq1$ and PER$_D\leq1$ are satisfied in the analysis
of the MSM and DSM, respectively,
the perturbativity bounds of Eq. (\ref{perb}) are satisfied in the respective
models. We describe the role
of PER$_M$ and PER$_D$ in our analysis later.

The Yukawa couplings $f^M$ and $f^D$ depend on neutrino masses and mixing angles,
which are found from the neutrino oscillation data.
From the global fits to neutrino oscillation
data, the following mass-square differences among the neutrino fields are
found \cite{deSalas:2020pgw}.
\begin{equation}
m_s^2=m_2^2-m_1^2=7.5\times10^{-5}~{\rm eV}^2,\quad
m_a^2=\left\{\begin{array}{c}
m_3^2-m_1^2=2.55\times10^{-3}~{\rm eV}^2~~{\rm (NO)}\\
m_1^2-m_3^2=2.45\times10^{-3}~{\rm eV}^2~~{\rm (IO)}
\end{array}\right..
\label{msq}
\end{equation}
Here, NO(IO) represents normal(inverted) ordering. In the above equation,
we have given the best fit values. In order to fit the
above mass-square differences, we take the neutrino mass eigenvalues as
\begin{eqnarray}
&& {\rm NO}:\quad m_1\lapprox m_s,\quad m_2=\sqrt{m_s^2+m_1^2},\quad
m_3=\sqrt{m_a^2+m_1^2}.
\nonumber \\
&& {\rm IO}:\quad m_3\lapprox m_s,\quad m_1=\sqrt{m_a^2+m_3^2},\quad
m_2=\sqrt{m_s^2+m_1^2}.
\label{nmass}
\end{eqnarray}
Notice that the lightest neutrino mass in NO and IO is still undetermined. Hence, in
our analysis, we have varied the lightest neutrino mass in the range $[0,m_s]$.
Also, in our analysis, we have applied the cosmological upper bound
on the sum of neutrino masses, which is 0.12 eV \cite{Planck:2018vyg}. Apart
from the neutrino masses, neutrino mixing angles and $CP$ violating Dirac phase
are also found from the global fits to neutrino oscillation
data \cite{deSalas:2020pgw}. The best fit and 3$\sigma$ ranges for these variables
are given in Tab. \ref{t4}.
\begin{table}[!h]
\centering
\begin{tabular}{|c|c c|} \hline
parameter & best fit & 3$\sigma$ range \\\hline
$\sin^2\theta_{12}/10^{-1}$ & 3.18 & 2.71 - 3.69 \\
$\sin^2\theta_{13}/10^{-2}$ (NO) & 2.200 & 2.000 - 2.405 \\
$\sin^2\theta_{13}/10^{-2}$ (IO) & 2.225 & 2.018 - 2.424 \\
$\sin^2\theta_{23}/10^{-1}$ (NO) & 5.74 & 4.34 - 6.10 \\
$\sin^2\theta_{23}/10^{-1}$ (IO) & 5.78 & 4.33 - 6.08 \\
$\delta_{CP}/{\rm o}$ (NO) & 194 & 128 - 359 \\
$\delta_{CP}/{\rm o}$ (IO) & 284 & 200 - 353 \\\hline
\end{tabular}
\caption{Best fit and 3$\sigma$ ranges for the neutrino mixing angles
and $CP$ violating Dirac phase, which are obtained from the global
fits to neutrino oscillation data \cite{deSalas:2020pgw}.}
\label{t4}
\end{table}
Notice that, in the case of MSM, the Yukawa couplings $f^M$ can depend on two Majorana
phases through $U_{PMNS}$. However, these Majorana phases are not determined
in experiments so far. Hence, we have taken these phases to be zero in our
analysis.

\subsection{Numerical results in the MSM}
\label{s6.1}

The numerical results on the LFV observables in the MSM are obtained after
scanning over the parameter space of this model. Below we describe our scanning
procedure of parameters in the MSM.
From Sec. \ref{s4}, notice that the analytical expressions
of all the LFV processes depend on the Yukawa couplings and the masses of the
fields $\eta_M^\pm$, $N_k^M$. The Yukawa couplings in our case, which are
given in Eq. (\ref{fm}), are determined by the neutrino oscillation observables
and the masses of the fields $\eta_{MR}$, $\eta_{MI}$, $N_k^M$. As a result of
this, numerical results of all the LFV observables in the MSM are obtained after
specifying the neutrino masses, neutrino mixing angles, $\delta_{CP}$ and the
masses of the fields $\eta_M^\pm$, $\eta_{MR}$, $\eta_{MI}$, $N_k^M$. Neutrino
masses, in our analysis, are computed and are varied according to the
description given around
Eq. (\ref{nmass}). As for the neutrino mixing angles and $\delta_{CP}$, we
have varied them over the 3$\sigma$ ranges given in Tab. \ref{t4}.
Now, the masses of the fields
$\eta_M^\pm$, $\eta_{MR}$, $\eta_{MI}$ and $N_k^M$ can be expressed in terms of
$m^2_{\eta_M^\pm}$, $\lambda_4$, $\lambda_5$ and $M_k$. In our analysis, we have
taken $N_k^M$ to be non-degenerate and parameterized their masses as $M_1$,
$M_1+\delta_M$, $M_1+2\delta_M$. Here, $\delta_M$ gives mass splitting between two
$N_k^M$ fields. The parameter $\lambda_4$ is varied in the range $[-4\pi,4\pi]$.
Similarly, the parameter $\lambda_5$ is varied over a range, which we mention
later in this section. In our analysis, we have done multiple scans over the parameter
space of the MSM. In all these scans, we have fixed $\delta_M$ to a specific
value. On the other hand, the values of $m^2_{\eta_M^\pm}$ and $M_1$ are either
fixed to a specific value or varied over a range, which we mention later.
However, we have varied the above described parameters in such a way that the
following bounds are satisfied on the masses of $\eta_{MR}$ and $\eta_{MI}$:
\begin{equation}
m_R\geq5~{\rm GeV},\quad m_I\geq5~{\rm GeV}
\end{equation}

\begin{figure}[!h]
\centering

\includegraphics[width=3.0in]{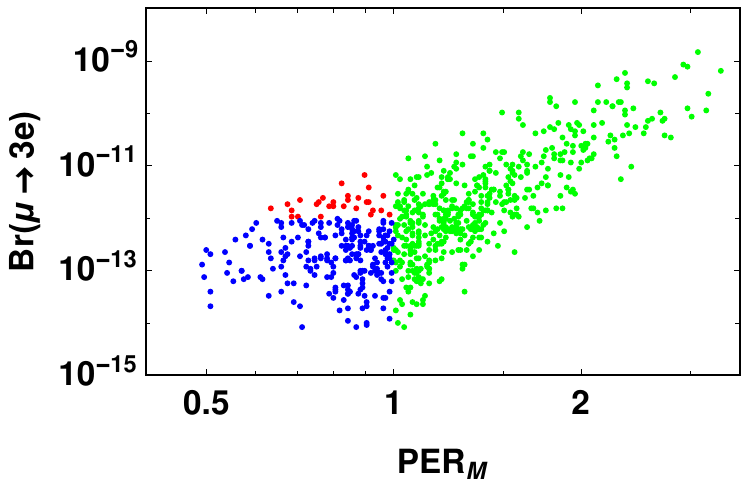}
\includegraphics[width=3.0in]{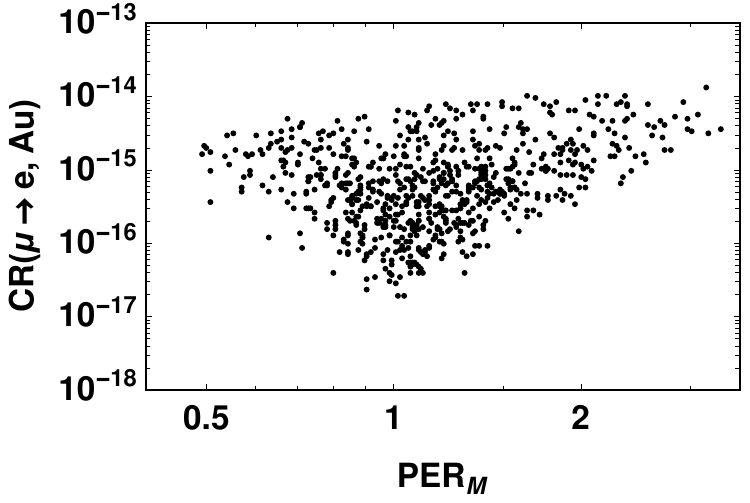}

\caption{${\rm Br}(\mu\to3e)$ and ${\rm CR}(\mu\to e,{\rm Au})$ versus PER$_M$,
after satisfying the experimental bound on ${\rm Br}(\mu\to e\gamma)$. The plots are
for the case of NO and $m_{\eta_M^\pm}<M_1$. The color coding in the left-hand side
plot is as follows:
green points violate the perturbativity bound on the Yukawa couplings, red points violate
the experimental bound on ${\rm Br}(\mu\to3e)$, blue points are allowed by the
above mentioned bounds. See text, for more details.}
\label{f3}
\end{figure}

Above, we have described our scanning procedure of parameters in the MSM, through
which we compute all the LFV observables. Now,
we first present results on the following LFV observables: ${\rm Br}(\mu\to e\gamma)$,
${\rm Br}(\mu\to3e)$, ${\rm CR}(\mu\to e,{\rm Au})$. Among these observables,
the experimental limit on ${\rm Br}(\mu\to e\gamma)$ is more stringent. Hence, in
our analysis, after satisfying the experimental bound on ${\rm Br}(\mu\to e\gamma)$,
we have computed the quantities ${\rm Br}(\mu\to3e)$ and ${\rm CR}(\mu\to e,{\rm Au})$.
These results are presented in Fig. \ref{f3} for the case of NO.
In this figure, we have fixed $m_{\eta_M^\pm}=$ 500 GeV and $\delta_M=$ 1 TeV.
We have varied $M_1$ and $\lambda_5$ in the ranges $[10,100]$ TeV and
$[10^{-11},10^{-10}]$, respectively. From the left-hand side plot of Fig.
\ref{f3}, we see that the blue points are allowed from the constraints of both
perturbativity of Yukawa couplings and experimental bound of ${\rm Br}(\mu\to3e)$.
The green points in this plot are disallowed due to violating the perturbativity bounds
on the Yukawa couplings. Notice that some of these green points also violate the
experimental bound on ${\rm Br}(\mu\to3e)$. On the other hand, the red points
in the above mentioned plot, satisfy the perturbativity bound on Yukawa couplings
but violate the experimental bound on ${\rm Br}(\mu\to3e)$. From the right-hand
side plot of Fig. \ref{f3}, we see that the values of ${\rm CR}(\mu\to e,{\rm Au})$
are within the experimental bound on this observable, even if the perturbativity
bound on the Yukawa couplings is violated. What this implies is that if the
experimental bound on ${\rm Br}(\mu\to e\gamma)$ is satisfied, the observable
${\rm CR}(\mu\to e,{\rm Au})$ may not give further constraints on the parametric space
of the model. On the other hand, ${\rm Br}(\mu\to3e)$ can impose additional
constraints on the parametric space. Notice that, in addition to ${\rm Br}(\mu\to3e)$,
PER$_M$ gives a significant constraint on the parametric space, which is evident from
the left-hand side plot of Fig. \ref{f3}.

Plots in Fig. \ref{f3} are illustrative to show the role of PER$_M$,
${\rm Br}(\mu\to3e)$ and ${\rm CR}(\mu\to e,{\rm Au})$ in constraining the parametric
space of the model, in addition to the constraints imposed by ${\rm Br}(\mu\to e\gamma)$.
In this figure, we have taken specific values for $m_{\eta_M^\pm}$ and $\delta_M$,
and specific ranges for $M_1$ and $\lambda_5$.
Below we describe what happens if we vary one of the
above quantities and keeping the others same as in Fig. \ref{f3}.
We have varied $m_{\eta_M^\pm}$ to 1 TeV in the plots of Fig. \ref{f3} and we have
found the plots remain qualitatively same. Similarly, after varying $\delta_M$, the
plots in Fig. \ref{f3} have remained qualitatively same. The values of $M_1$ in
Fig. \ref{f3} are large compared to that of $m_{\eta_M^\pm}$. If we decrease $M_1$
less than 10 TeV in Fig. \ref{f3}, we have found PER$_M$ to be less than 1. In this
case, we get very few number of points which violate the experimental bound on
${\rm Br}(\mu\to3e)$ in addition to the points which satisfy this experimental bound.
On the other hand, by increasing $M_1$ more than 100 TeV in
Fig. \ref{f3}, PER$_M$ is found to be larger than 1. This case can be completely
ruled out by solely imposing constraints of ${\rm Br}(\mu\to e\gamma)$ and PER$_M$.
Lastly, by varying $\lambda_5$ to be below $10^{-11}$ in Fig. \ref{f3}, PER$_M$ is to found
to be greater than 1. Hence, this case is completely ruled out with the constraints
of ${\rm Br}(\mu\to e\gamma)$ and PER$_M$. Now, by varying $\lambda_5$ to be above
$10^{-10}$ in Fig. \ref{f3}, we have found PER$_M$ to be less than 1 and all the points
would satisfy the constraint on ${\rm Br}(\mu\to3e)$.
In all the above described cases, we have found the values of ${\rm CR}(\mu\to e,{\rm Au})$
to be below the experimental bound on this observable.

Notice that PER$_M$ is inversely related to $\Lambda_k^M$, which in turn is inversely
related to either $m_{\eta_M^\pm}$ or $M_k$. Hence, PER$_M$ increases with $M_k$.
On the other hand, $\Lambda_k^M\to 0$ in the limit $\lambda_5\to0$. Hence, by
decreasing $\lambda_5$, PER$_M$ increases. The above described features supplement
the description given in the previous paragraph.

From the plots of Fig. \ref{f3}, we have noticed that, among the different LFV
observables, ${\rm Br}(\mu\to e\gamma)$ and ${\rm Br}(\mu\to3e)$ can
impose constraints on the parametric space of the model, if not by
${\rm CR}(\mu\to e,{\rm Au})$. We can understand this in the following way.
Both the processes $\mu\to e\gamma$ and $\mu{\rm Au}\to e{\rm Au}$ are driven
by the diagrams of Fig. \ref{f1}. Hence, after making naive order of estimation
in the expressions of ${\rm Br}(\mu\to e\gamma)$ and ${\rm CR}(\mu\to e,{\rm Au})$,
one can arrive at ${\rm CR}(\mu\to e,{\rm Au})<{\rm Br}(\mu\to e\gamma)$ for any
parametric point in the model. Since the experimental bound on ${\rm Br}(\mu\to e\gamma)$
is stringent than that on ${\rm CR}(\mu\to e,{\rm Au})$, ${\rm CR}(\mu\to e,{\rm Au})$
may not give additional constraints on the model if the bound on
${\rm Br}(\mu\to e\gamma)$ is satisfied. On the other hand, for the process
$\mu\to3e$, box diagrams of Fig. \ref{f2} contribute, in addition to the diagrams
of Fig. \ref{f1}. Notice that the amplitudes of box diagrams are proportional to
quartic in Yukawa couplings, whereas, the amplitudes of diagrams of Fig. \ref{f1}
vary as quadratic in Yukawa couplings. Hence, for large values of Yukawa couplings,
the contribution of box diagrams can overcome that of Fig. \ref{f1}. Indeed, from
the left-hand side plot of Fig. \ref{f3}, we see that for PER$_M\gapprox1$,
${\rm Br}(\mu\to3e)$ can exceed over the experimental bound on it. We have noticed
that in the region of PER$_M\sim1$, contribution from box diagrams in
${\rm Br}(\mu\to3e)$ is dominant. Hence,
for large Yukawa couplings, ${\rm Br}(\mu\to3e)$ can give additional constraints
on the model parameters in addition to that from ${\rm Br}(\mu\to e\gamma)$.

\begin{figure}[!h]
\centering

\includegraphics[width=3.0in]{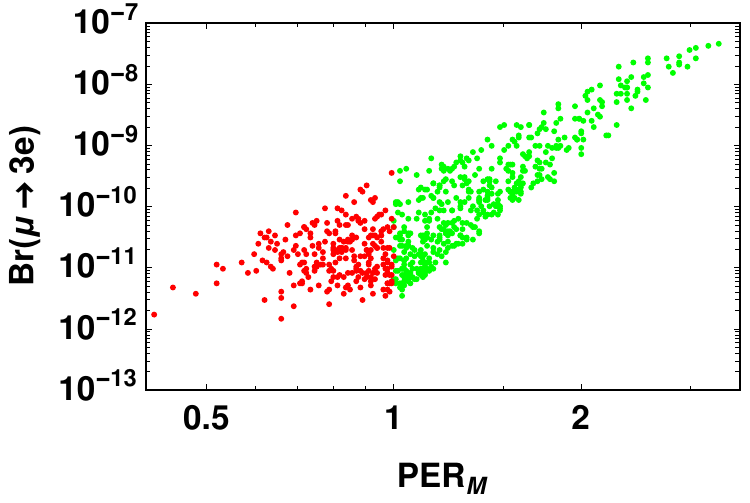}
\includegraphics[width=3.0in]{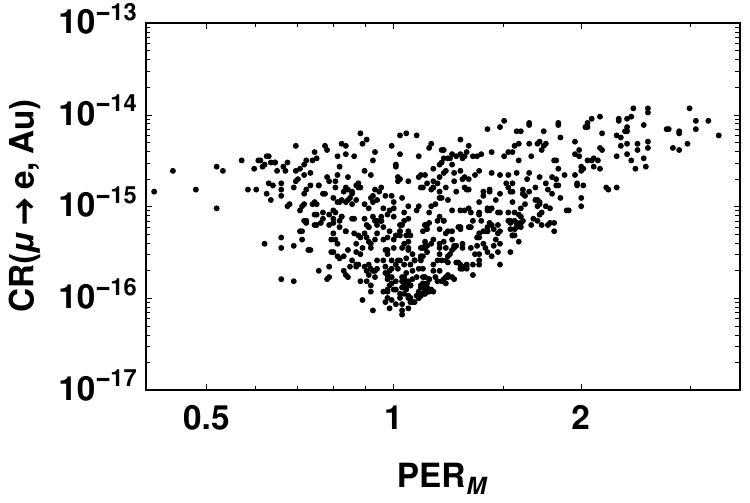}

\caption{${\rm Br}(\mu\to3e)$ and ${\rm CR}(\mu\to e,{\rm Au})$ versus PER$_M$,
after satisfying the experimental bound on ${\rm Br}(\mu\to e\gamma)$. The plots are
for the case of IO and $m_{\eta_M^\pm}<M_1$. The color coding in the left-hand side
plot is same as that
in Fig. \ref{f3}. See text, for more details.}
\label{f4}
\end{figure}

Results about ${\rm Br}(\mu\to3e)$ and ${\rm CR}(\mu\to e,{\rm Au})$, in the
case of IO, are presented in Fig. \ref{f4}, after satisfying the experimental bound
on ${\rm Br}(\mu\to e\gamma)$.
In this figure, the values of $m_{\eta_M^\pm}$ and $\delta_M$ are same as that
in Fig. \ref{f3}. Moreover, the ranges of $M_1$ and $\lambda_5$ are same as
that in Fig. \ref{f3}. From the left-hand side plot of Fig. \ref{f4}, we see
that the whole parametric region is ruled out either with the perturbativity
bound on Yukawa couplings or by the experimental bound on ${\rm Br}(\mu\to3e)$.
On the other hand, from the right-hand side of this figure, notice that
${\rm CR}(\mu\to e,{\rm Au})$ is not giving any additional constraints on the
parametric region. Plots in this figure are illustrative to show the role played
by ${\rm Br}(\mu\to3e)$ and PER$_M$ in giving constraints on the model parameters,
in addition to the constraints imposed by ${\rm Br}(\mu\to e\gamma)$. After comparing
the plots of this figure with that of Fig. \ref{f3}, notice that ${\rm Br}(\mu\to3e)$
and ${\rm CR}(\mu\to e,{\rm Au})$ are enhanced in the case of IO as compared to
that of NO. The reason for this enhancement is that, in the case of IO, the neutrino
mass eigenvalues are large as compared to that in NO. As a result of this,
the Yukawa couplings are large in the case of IO as compared to that of NO.

\begin{figure}[!h]
\centering

\includegraphics[width=3.0in]{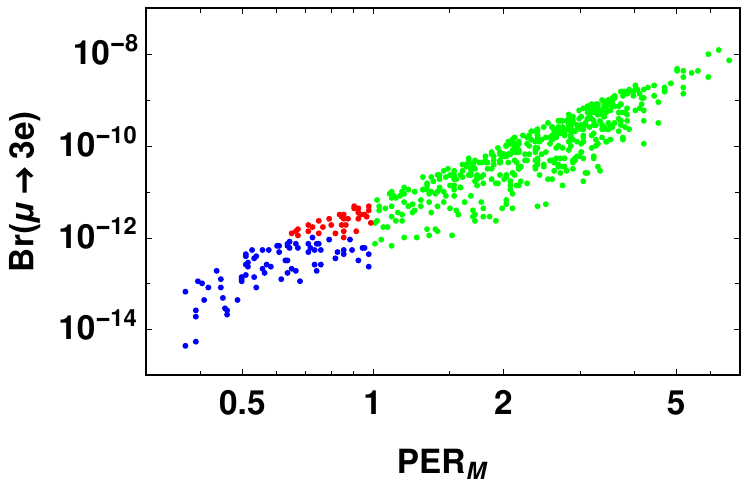}
\includegraphics[width=3.0in]{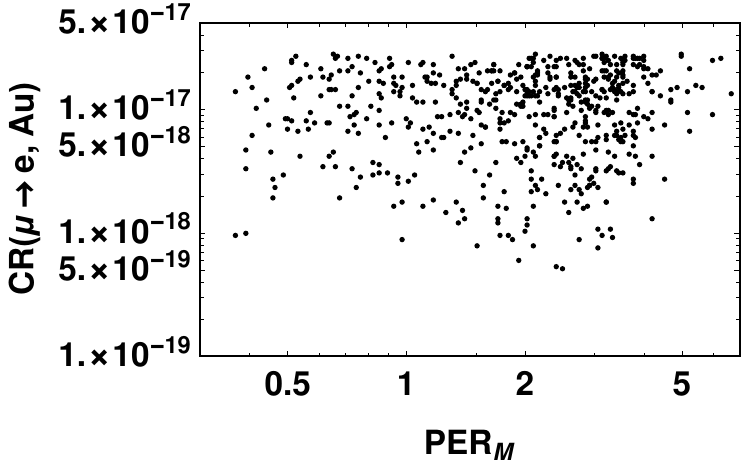}

\caption{${\rm Br}(\mu\to3e)$ and ${\rm CR}(\mu\to e,{\rm Au})$ versus PER$_M$,
after satisfying the experimental bound on ${\rm Br}(\mu\to e\gamma)$. The plots are
for the case of IO and $m_{\eta_M^\pm}>M_1$. The color coding in the left-hand side
plot is same as that
in Fig. \ref{f3}. See text, for more details.}
\label{f5}
\end{figure}

In Figs. \ref{f3} and \ref{f4}, we have explained
that ${\rm Br}(\mu\to3e)$ can give additional constraints
on the parametric space, even if PER$_M<1$ and
${\rm Br}(\mu\to e\gamma)<3.1\times10^{-13}$. However, these plots are made for the
case of $m_{\eta_M^\pm}<M_1$. We have done the above mentioned analysis for the
case $m_{\eta_M^\pm}>M_1$. Now, in this case and for NO, we have not found a parametric
space where ${\rm Br}(\mu\to3e)$ can give additional constraints if PER$_M<1$ and
${\rm Br}(\mu\to e\gamma)<3.1\times10^{-13}$. On the other hand, in the case of IO
and for $m_{\eta_M^\pm}>M_1$, we have found a region where ${\rm Br}(\mu\to3e)$ can
give additional constraints in addition to that of PER$_M$ and
${\rm Br}(\mu\to e\gamma)$. These results are presented in Fig. \ref{f5}.
In this figure we have taken $M_1=$ 1 TeV and $\delta_M=$ 500 GeV. We have varied
$m_{\eta_M^\pm}$ and $\lambda_5$ in the ranges $[10,100]$ TeV and $[10^{-9},10^{-8}]$,
respectively. From the left-hand side plot of this figure, we see that red points
indicate the region where ${\rm Br}(\mu\to3e)$ give additional constraints even
if PER$_M<1$ and ${\rm Br}(\mu\to e\gamma)<3.1\times10^{-13}$. From the right-hand
side plot of this figure, notice that ${\rm CR}(\mu\to e,{\rm Au})$ does not give
additional constraint.

\begin{figure}[!h]
\centering

\includegraphics[width=3.0in]{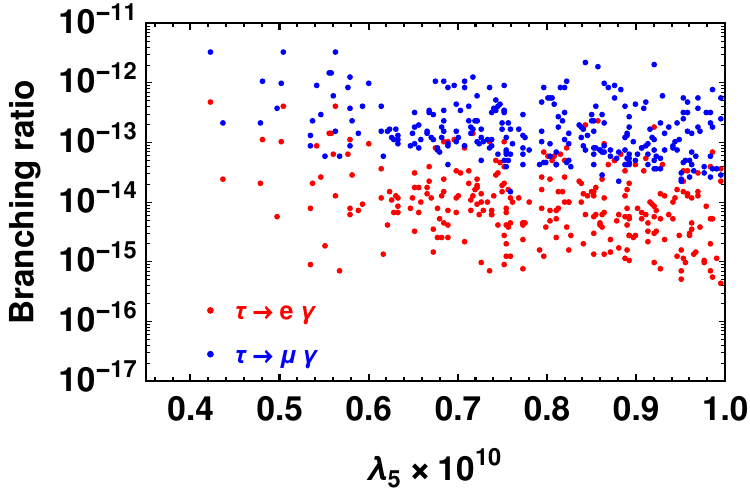}
\includegraphics[width=3.0in]{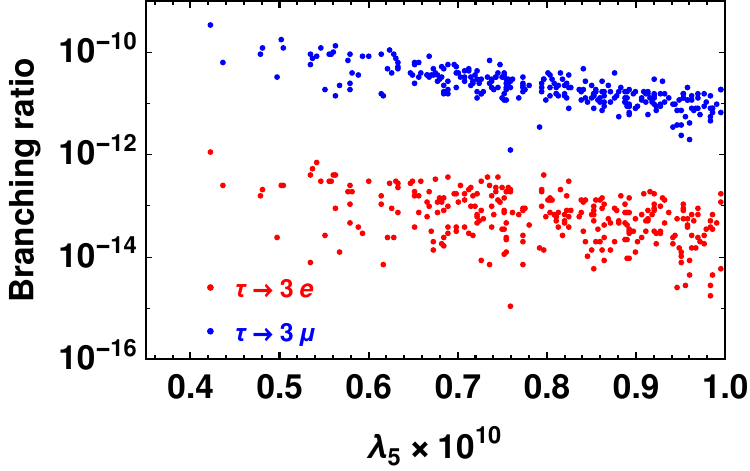}

\includegraphics[width=3.0in]{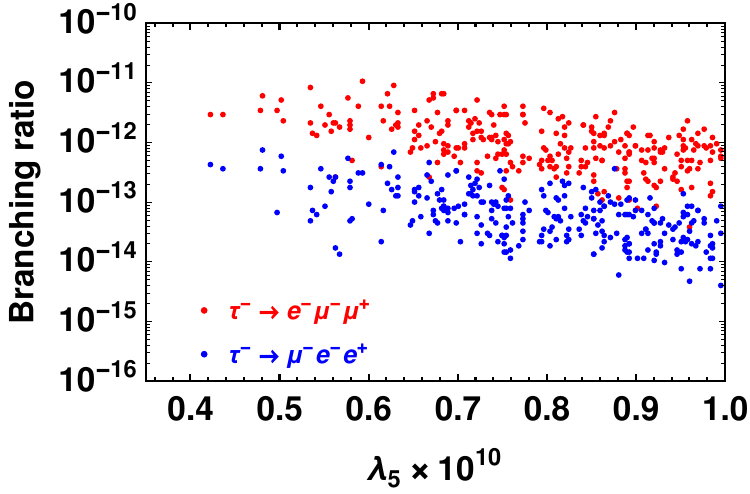}
\includegraphics[width=3.0in]{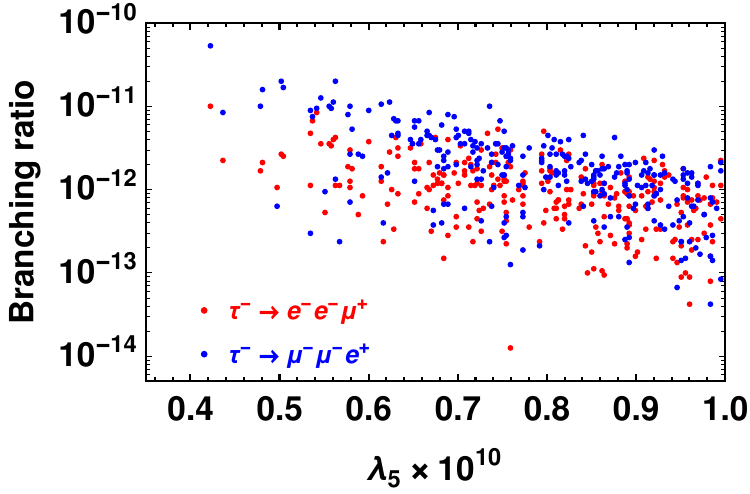}

\caption{Branching ratios of various LFV tau decays in the MSM and for the case of NO,
after applying
the constraints due to PER$_M$, ${\rm Br}(\mu\to e\gamma)$, ${\rm Br}(\mu\to3e)$
and ${\rm CR}(\mu\to e,{\rm Au})$. See text, for more details.}
\label{f6}
\end{figure}

In Figs. \ref{f3}$-$\ref{f5}, we have shown how parameter space of the MSM can be
constrained by PER$_M$, ${\rm Br}(\mu\to e\gamma)$, ${\rm Br}(\mu\to3e)$
and ${\rm CR}(\mu\to e,{\rm Au})$. By applying the constraints due to above
mentioned quantities, we have obtained branching ratios of the LFV tau decays.
These results are presented in Fig. \ref{f6} for the case of NO.
Notice that,
we have described before that ${\rm CR}(\mu\to e,{\rm Au})$ is not giving
additional constraints on the model parameters. However, for the sake of completeness
we have satisfied the experimental upper bound on ${\rm CR}(\mu\to e,{\rm Au})$
in Fig. \ref{f6}.
In this figure, we have fixed $\delta_M=$ 1 TeV. We have varied $m_{\eta_M^\pm}$ and
$M_1$ in the ranges $[100,1000]$ GeV and $[5,50]$ TeV, respectively, in this
figure. For the above mentioned ranges, we have found that the branching ratios
in this figure to be maximal. Notice that ${\rm Br}(\tau\to3\mu)$ can reach the
maximum as around $10^{-10}$. This value is two orders less than the current
experimental bound on ${\rm Br}(\tau\to3\mu)$. In future experiments, this decay
branching ratio will be probed to a sensitivity of ${\cal O}(10^{-10})$
\cite{Belle-II:2018jsg,FCC:2018byv}. Hence, the MSM can be tested in the
future experiments in the decay channel $\tau\to3\mu$. The other LFV tau decays
in the MSM can reach the maximum as around $10^{-13}-10^{-11}$. These values are
three to four orders less than the current experimental bounds on these decays.
In Fig. \ref{f6}, notice that $\lambda_5$ should be $\gapprox 4\times10^{-11}$.
In our scan, the value of $\lambda_5$ is restricted by the constraints due to
PER$_M$, ${\rm Br}(\mu\to e\gamma)$, ${\rm Br}(\mu\to3e)$ and
${\rm CR}(\mu\to e,{\rm Au})$. Indeed, in our scan, if we vary $m_{\eta_M^\pm}$
and $M_1$ independently over the range $[100,10^5]$ GeV, we have found $\lambda_5$
is restricted
to be $\gapprox \times10^{-11}$. Since the values of $\lambda_5$ given in
Fig. \ref{f6} are the lowest allowed values, we see that the Yukawa couplings
should be large for these values, whose statement is explained in Sec. \ref{s2.1}.
As a result of the above discussion, we expect the branching ratios of LFV tau decays
to be maximal. In fact, this statement is vindicated in our analysis, where we have
found that, when PER$_M\sim1$ the branching ratios of LFV tau decays are becoming
maximal. 

\begin{figure}[!h]
\centering

\includegraphics[width=3.0in]{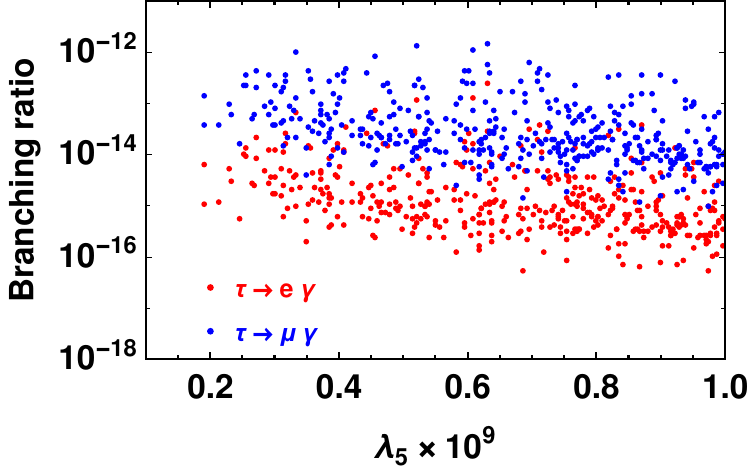}
\includegraphics[width=3.0in]{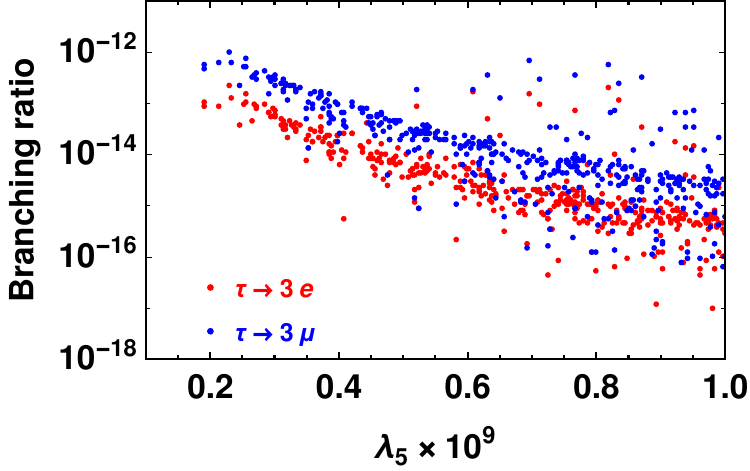}

\includegraphics[width=3.0in]{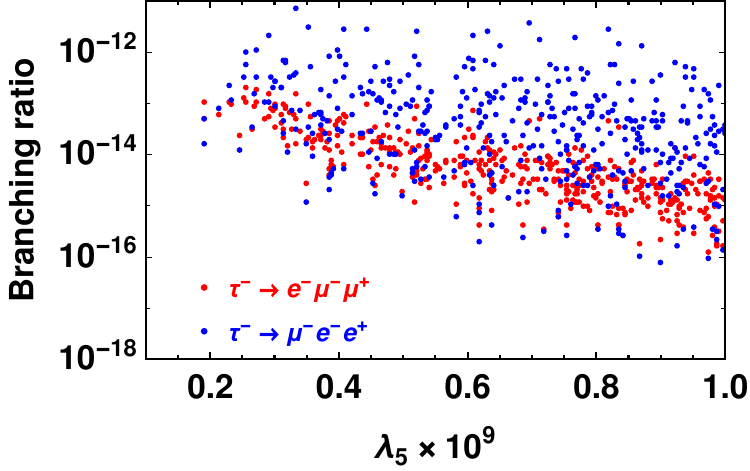}
\includegraphics[width=3.0in]{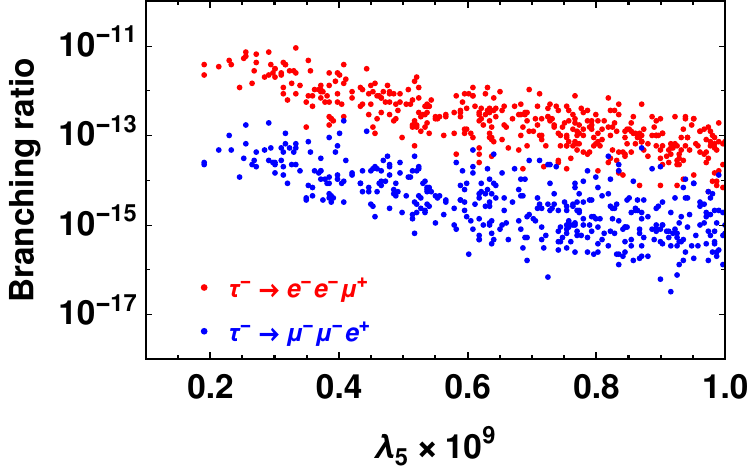}

\caption{Branching ratios of various LFV tau decays in the MSM and for the case of IO,
after applying
the constraints due to PER$_M$, ${\rm Br}(\mu\to e\gamma)$, ${\rm Br}(\mu\to3e)$
and ${\rm CR}(\mu\to e,{\rm Au})$. See text, for more details.}
\label{f7}
\end{figure}

We have obtained analog plots of Fig. \ref{f6}, for the case of IO. These results
are presented in Fig. \ref{f7}.
In this figure, we have fixed $\delta_M=$ 1 TeV. We have varied $m_{\eta_M^\pm}$ and
$M_1$ independently in the range $[.5,50]$ TeV. For the above mentioned range, we have found
the branching ratios in this figure to be maximal. From this figure, we notice
that the branching ratio of $\tau^-\to e^-e^-\mu^+$ can reach as around $10^{-11}$. The
branching ratio of other LFV tau decays can reach as around $10^{-13}-10^{-12}$. Notice
that the maximal values of branching ratios of the decays $\tau\to e\gamma$,
$\tau\to\mu\gamma$, $\tau^-\to\mu^-e^-e^+$ and $\tau^-\to e^-e^-\mu^+$
in Fig. \ref{f7} are nearly same as that of the corresponding decays in Fig. \ref{f6}.
On the other hand, for the rest of the
LFV tau decays, the maximal values of the branching ratios in Fig. \ref{f7} are
suppressed as compared to that of the corresponding decays in Fig. \ref{f6}.
This implies that we can test the
LFV tau decays in future experiments in the case of NO rather than that for
the case of IO. The value of $\lambda_5$ in
Fig. \ref{f7} is restricted by the constraints due to PER$_M$,
${\rm Br}(\mu\to e\gamma)$, ${\rm Br}(\mu\to3e)$ and ${\rm CR}(\mu\to e,{\rm Au})$.
From this figure, we see that the $\lambda_5$ should be $\gapprox10^{-10}$. Notice
that the allowed value of $\lambda_5$, in the case of IO, is larger than that
for the case of NO. As a result of this, we expect the magnitude of allowed Yukawa
couplings in the case of IO to be lower than that for the case of NO.
This may be a reason for why the
maximal values of branching ratios of LFV tau decays in the case of IO are unable
to exceed than that for the case of NO.

\subsection{Numerical results in the DSM}

Before presenting the numerical results in the DSM, below we describe our scanning procedure
through which we determine all the LFV observables of this model. The LFV observables,
after using the analytical expressions of Sec. \ref{s4}, depend on the Yukawa couplings
and the masses of the fields $\eta_D^\pm$, $N_k^D$. Notice that the
Yukawa couplings, which are given in Eq. (\ref{fd}), depend on the neutrino
oscillation observables,
masses of the fields $N_k^D$, $\zeta_{1,2}$ and the mixing parameter $\theta$.
Hence, all the LFV observables of the DSM are determined by neutrino oscillation
observables, masses of the fields $\eta_D^\pm$, $N_k^D$, $\zeta_{1,2}$ and the
mixing parameter $\theta$.
As for the neutrino oscillation observables, we vary them according to the
description given in Sec. \ref{s6.1}.
Now, the masses of the fields $\eta_D^\pm$, $N_k^D$, $\zeta_{1,2}$ and the
mixing parameter $\theta$ can be parameterized in terms of the variables
$m_{\eta_D^\pm}$, $M_k^\prime$, $m_\chi$, $\lambda_4^\prime$, $\lambda_5^\prime$
and $\frac{A}{v}$. Here, $M_k^\prime$ indicate the mass of the fields $N_k^D$,
which we have taken to be non-degenerate in our analysis. We have parameterized
the masses of $N_k^D$ as $M_1^\prime$, $M_1^\prime+\delta_M^\prime$ and
$M_1^\prime+2\delta_M^\prime$, where $\delta_M^\prime$ gives splitting in the
masses. The parameters $\lambda_{4,5}^\prime$ are varied independently
over $[-4\pi,4\pi]$. In analogy to the results preseneted in Sec. \ref{s6.1},
in the DSM we have done multiple scans over the parameter space. In all these
scans, $\delta_M^\prime$ has been fixed to a specific value. On the other hand,
$m_{\eta_D^\pm}$ and $M_1^\prime$ are either fixed to some specific values
or varied over some particular ranges. Whereas, the parameters $m_\chi$ and
$\frac{A}{v}$ are varied over some ranges. Later in our analysis, we specify
the values taken for the above mentioned variables. However, we have varied them in
such a way that the following lower bounds are satisfied on the masses of
$\zeta_{1,2}$:
\begin{equation}
m_{\zeta_1}\geq 5~{\rm GeV},\quad m_{\zeta_2}\geq 5~{\rm GeV}
\end{equation}

\begin{figure}[!h]
\centering

\includegraphics[width=3.0in]{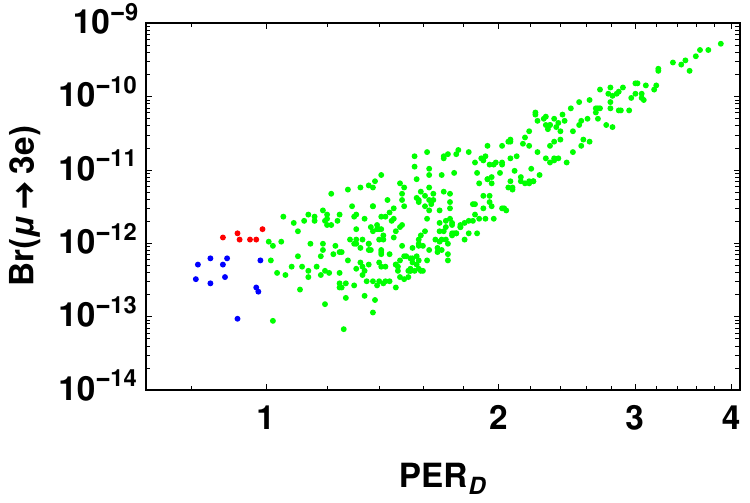}
\includegraphics[width=3.0in]{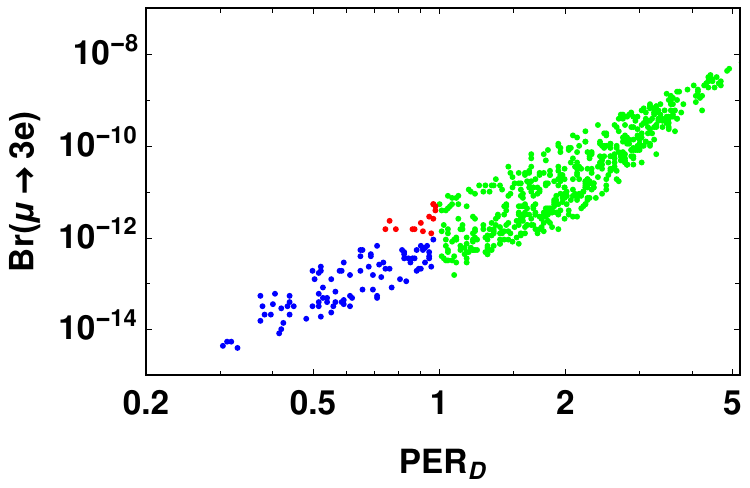}
\includegraphics[width=3.0in]{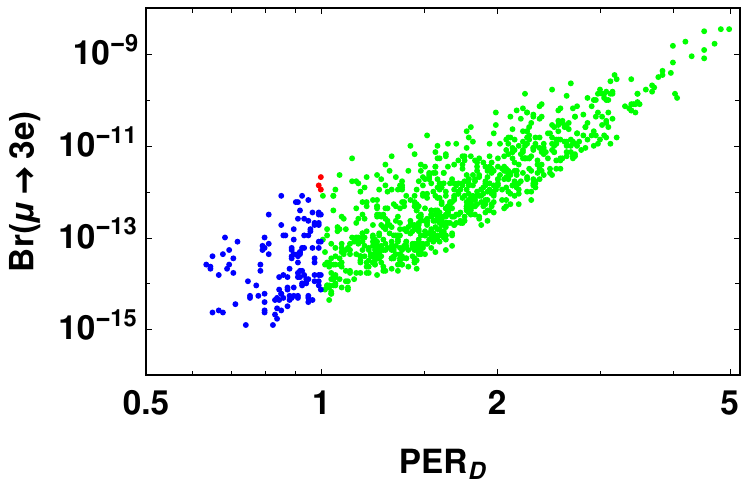}

\caption{${\rm Br}(\mu\to3e)$ versus PER$_D$,
after satisfying the experimental bound on ${\rm Br}(\mu\to e\gamma)$. The plots are
for the case of IO. Top-left and -right plots are for the cases $m_{\eta_D^\pm}<M_1^\prime$
and $m_{\eta_D^\pm}>M_1^\prime$, respectively. The bottom plot is for the case
$m_{\eta_D^\pm}\sim M_1^\prime$. The color coding in these plots is same as that
in Fig. \ref{f3}. See text, for more details.}
\label{f8}
\end{figure}

Above, we have described our scanning procedure of parameters in the DSM.
Now, in analogy to the results presented in the previous subsection, here
we first present results on the following LFV observables: ${\rm Br}(\mu\to e\gamma)$,
${\rm Br}(\mu\to3e)$, ${\rm CR}(\mu\to e,{\rm Au})$.
In our scanning procedure, we have computed the observables ${\rm Br}(\mu\to3e)$
and ${\rm CR}(\mu\to e,{\rm Au})$ after satisfying the experimental bound on
${\rm Br}(\mu\to e\gamma)$. In our analysis we have found that
${\rm CR}(\mu\to e,{\rm Au})$ does not give additional constraints on the parameter
space of the DSM. This result is analogous to that presented in the previous
subsection for the case of MSM. The above result should not be surprising, since
both DSM and MSM are similar to each other in terms of model perspective.
Next, we have searched
for ${\rm Br}(\mu\to3e)$ to give any additional constraints on the model
parameters, after satisfying the experimental bound on ${\rm Br}(\mu\to e\gamma)$
and also the perturbativity bound on the Yukawa couplings. The perturbativity
bound of Yukawa couplings in the DSM is satisfied by imposing PER$_D<1$. In our
scanning analysis, in the case of NO, after imposing
${\rm Br}(\mu\to e\gamma)<3.1\times10^{-13}$ and PER$_D<1$, we have found that
${\rm Br}(\mu\to3e)$ does not give additional constraints on the parameter
space of the model. On the other hand, in the case of IO, ${\rm Br}(\mu\to3e)$
is found to give additional constraints on the model parameters. These
results are presented in Fig. \ref{f8}.
The plots in this figure are illustrative, which are shown for various cases that
differ in the values of $m_{\eta_D^\pm}$ and $M_1^\prime$. In the top-left plot
of Fig. \ref{f8}, we have fixed $m_{\eta_D^\pm}=$ 500 GeV and $\delta_M^\prime=$
1 TeV. In this plot, we have varied $M_1^\prime$, $m_\chi$ and $\frac{A}{v}$
in the ranges $[10^4,10^5]$ GeV, $[10,10^5]$ GeV and $[10^{-11},10^{-10}]$,
respectively. Notice that this plot refers to the case of $m_{\eta_D^\pm}<M_1^\prime$.
In the top-right plot of Fig. \ref{f8}, which corresponds to the case
$m_{\eta_D^\pm}>M_1^\prime$, we have fixed $M_1^\prime=$ 500 GeV and
$\delta_M^\prime=$ 50 GeV. In this plot, we have varied $m_{\eta_D^\pm}$,
$m_\chi$ and $\frac{A}{v}$ in the ranges $[10^4,10^5]$ GeV, $[10,10^5]$ GeV and
$[10^{-9},10^{-8}]$, respectively. Finally, in the bottom plot of Fig. \ref{f8},
which represents the case $m_{\eta_D^\pm}\sim M_1^\prime$, we have fixed
$\delta_M^\prime=$ 1 TeV and varied $m_{\eta_D^\pm}$, $M_1^\prime$,
$m_\chi$ and $\frac{A}{v}$ in the ranges $[10^4,10^5]$ GeV, $[10^4,10^5]$ GeV,
$[10^3,10^4]$ GeV and $[10^{-10},10^{-9}]$, respectively. From the plots of
Fig. \ref{f8}, we see red points which indicate ${\rm Br}(\mu\to3e)>10^{-12}$
and PER$_D<1$. Essentially, these red points correspond to additional
constraints due to ${\rm Br}(\mu\to3e)$, even after satisfying the constraints
on ${\rm Br}(\mu\to e\gamma)$ and PER$_D$. Notice that there are only few
red points in Fig. \ref{f8} as compared to that in Figs. \ref{f3}$-$\ref{f5},
which are for the case of MSM. This implies that the additional constraints
due to ${\rm Br}(\mu\to3e)$ are not so severe in the DSM as compared to that
in the MSM. The reason for this can be understood in the following way. As
described in Sec. \ref{s6.1}, the constraints due to ${\rm Br}(\mu\to3e)$
become severe in the limit that the Yukawa couplings becoming large. It is
further explained that the box diagrams of Fig. \ref{f2} give dominant
contribution to ${\rm Br}(\mu\to3e)$ in the above mentioned limit. Now, it is
evident that both the box diagrams of Fig. \ref{f2} contribute to the decay
$\mu\to3e$ in the MSM. Whereas, only the left-hand side diagram of Fig. \ref{f2}
contribute to $\mu\to3e$ in the DSM. Since more contribution due to box diagrams
is possible in the MSM as against to that in the DSM, constraints due to
${\rm Br}(\mu\to3e)$ are more severe in the MSM as compared to that in the DSM.

\begin{figure}[!h]
\centering

\includegraphics[width=3.0in]{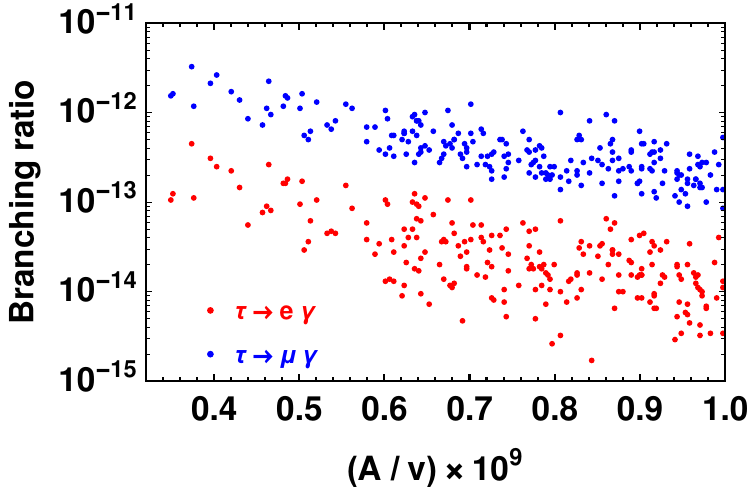}
\includegraphics[width=3.0in]{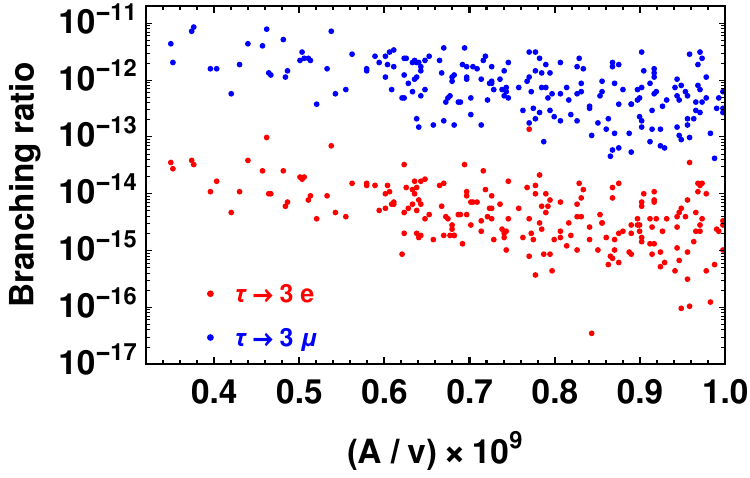}

\includegraphics[width=3.0in]{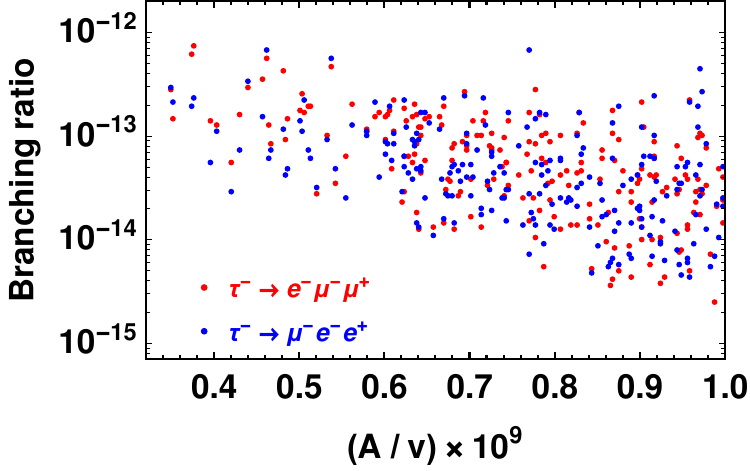}
\includegraphics[width=3.0in]{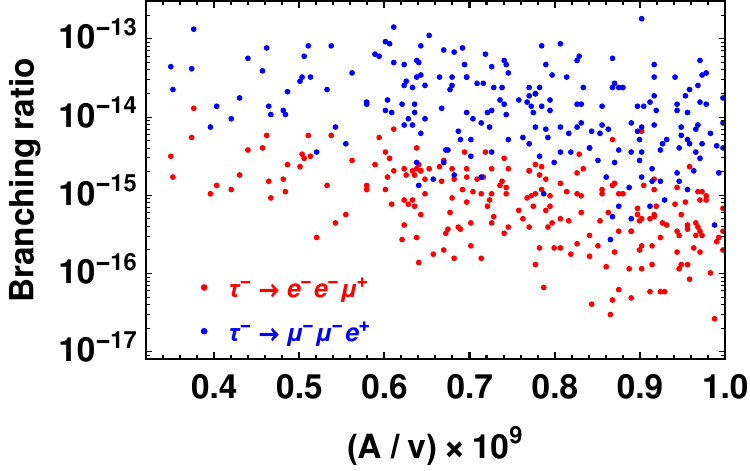}

\caption{Branching ratios of various LFV tau decays in the DSM and for the case of NO,
after applying
the constraints due to PER$_D$, ${\rm Br}(\mu\to e\gamma)$, ${\rm Br}(\mu\to3e)$
and ${\rm CR}(\mu\to e,{\rm Au})$. See text, for more details.}
\label{f9}
\end{figure}

In the previous paragraph, we have explained the constraints arising due to
PER$_D$, ${\rm Br}(\mu\to e\gamma)$, ${\rm Br}(\mu\to3e)$ and
${\rm CR}(\mu\to e,{\rm Au})$. After satisfying the constraints due to above
mentioned quantities, we have computed the branching ratios of LFV tau
decays in the DSM. These results are presented in Fig. \ref{f9}.
Although we have described before that ${\rm CR}(\mu\to e,{\rm Au})$ is not
giving additional constraints on the model parameters, for the sake of completeness
we have satisfied the experimental limit on this observable in Fig. \ref{f9}. In
this figure, we have fixed $\delta_M^\prime=$ 1 TeV and we have varied $m_{\eta_D^\pm}$,
$M_1^\prime$ and $m_\chi$ in the ranges $[5,50]$ TeV, $[5,50]$ GeV and $[50,500]$
GeV, respectively. For the above mentioned ranges, we have found that the branching
ratios in this figure can reach maximum values. Among the various LFV tau decays,
${\rm Br}(\tau\to3\mu)$ can have a maximum value of around $10^{-11}$, whose
value is an order less than the sensitivity to probe this decay channel
in the future experiments \cite{Belle-II:2018jsg,FCC:2018byv}.
The next best after $\tau\to3\mu$ is the decay $\tau\to\mu\gamma$, whose
branching ratio can reach a maximum of $\sim10^{-12}$. Rest of the branching
ratios of tau decays in Fig. \ref{f9} are further suppressed to be below $10^{-12}$.
Notice that the value of $\frac{A}{v}$ in Fig. \ref{f9} is restricted to be
$\gapprox3.2\times10^{-10}$. This restriction is due to the constraints of
PER$_D$, ${\rm Br}(\mu\to e\gamma)$, ${\rm Br}(\mu\to3e)$ and
${\rm CR}(\mu\to e,{\rm Au})$. In fact, in our scanning analysis, we have found
that $\frac{A}{v}$ is restricted to be larger than $10^{-10}$ even if we vary
$m_{\eta_D^\pm}$, $M_1^\prime$ and $m_\chi$ independently over a larger range of
$[100,10^5]$ GeV.
Since the values of $\frac{A}{v}$ are the lowest possible in Fig. \ref{f9},
the Yukawa couplings become large for these values, whose statement is explained in
Sec. \ref{s2.2}. As a result of this, we expect the branching ratios in Fig.
\ref{f9} to become maximal. Explicitly, in our analysis, we have seen that the
branching ratios in this figure reach maximum when PER$_D\sim1$.

Below we compare the results of Fig. \ref{f9} with the analog results in the case
of MSM, which
are presented in Fig. \ref{f7}. Notice that the maximal reaches of
${\rm Br}(\tau\to e\gamma)$ and ${\rm Br}(\tau\to \mu\gamma)$ are nearly same
in the DSM and MSM. On the other hand, for the rest of the LFV tau decays, which
are 3-body decays, the
maximal reaches of the branching ratios in the DSM are at least suppressed by an
order one as compared to that in the MSM. For these 3-body LFV tau decays,
the contribution
comes from the penguin diagrams of Fig. \ref{f1} and also from the box diagrams
of Fig. \ref{f2}. As already explained before, the branching ratios of these
decays become maximal when the Yukawa couplings are large, which implies
that the contribution
of box diagrams overcome that of dipole diagrams in these decays. Now, we
know that two different box diagrams contribute for the above decays in the MSM
as against to only one in the case of DSM. As a result of this, the maximal reaches
in the branching ratios of the 3-body LFV tau decays are suppressed in the DSM
as compared to that in the MSM. Now, as for the case of 2-body LFV decays
$\tau\to e\gamma$ and $\tau\to\mu\gamma$, the contribution to these decays come
only due to the penguin diagrams. Since the topology of penguin diagrams is
same between the MSM and DSM, the maximal reaches in the branching ratios of the
2-body LFV tau decays remain to be nearly same in the MSM and DSM.

\begin{figure}[!h]
\centering

\includegraphics[width=3.0in]{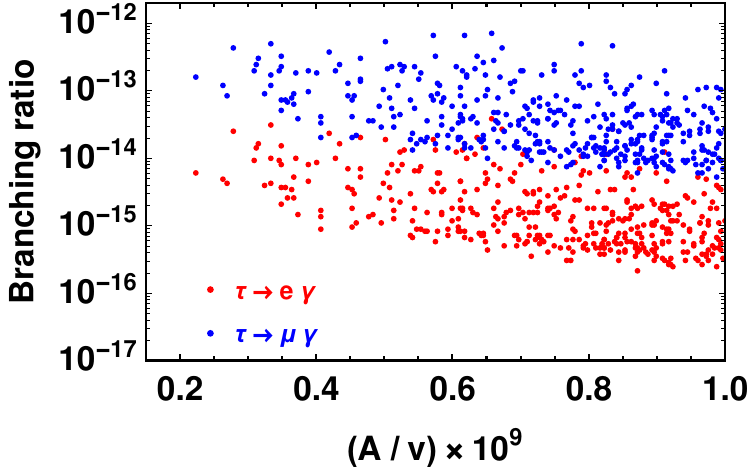}
\includegraphics[width=3.0in]{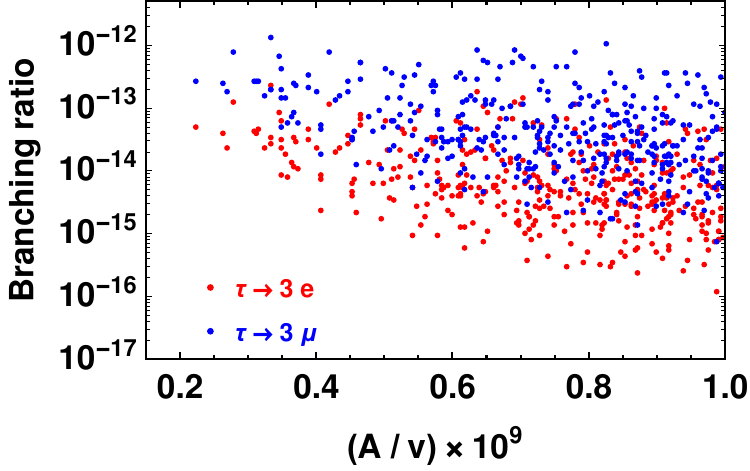}

\includegraphics[width=3.0in]{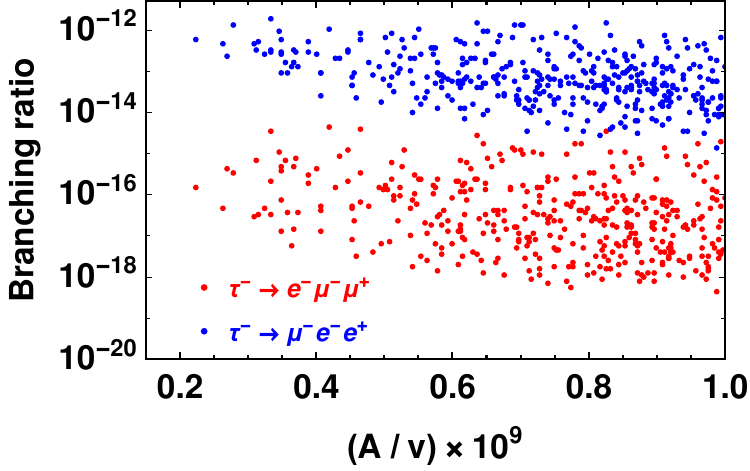}
\includegraphics[width=3.0in]{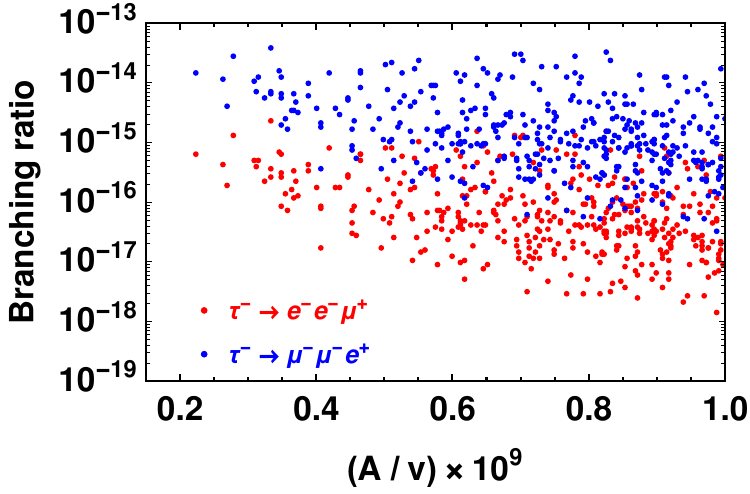}

\caption{Branching ratios of various LFV tau decays in the DSM and for the case of IO,
after applying
the constraints due to PER$_D$, ${\rm Br}(\mu\to e\gamma)$, ${\rm Br}(\mu\to3e)$
and ${\rm CR}(\mu\to e,{\rm Au})$. See text, for more details.}
\label{f10}
\end{figure}

We have obtained the analog plots of Fig. \ref{f9} for the case of IO. These
results are presented in Fig. \ref{f10}.
In this figure, we have fixed $\delta_M^\prime=$ 1 TeV and we have varied $m_{\eta_D^\pm}$,
$M_1^\prime$ and $m_\chi$ in the ranges $[5,50]$ TeV, $[5,50]$ TeV and $[50,500]$
GeV, respectively. For the above mentioned ranges, we have found that the branching
ratios in this figure can reach maximum values. In this figure, the value of
$\frac{A}{v}$ is restricted due to the constraints of PER$_D$, ${\rm Br}(\mu\to e\gamma)$,
${\rm Br}(\mu\to3e)$ and ${\rm CR}(\mu\to e,{\rm Au})$. The lowest allowed value
of $\frac{A}{v}$ in this figure is nearly the same as that in Fig. \ref{f9}.
Notice that ${\rm Br}(\tau\to\mu\gamma)$, ${\rm Br}(\tau\to3\mu)$ and
${\rm Br}(\tau^-\to\mu^-e^-e^+)$ can reach maximum values of around $10^{-12}$ in
Fig. \ref{f10}. For the rest of the LFV tau decays in Fig. \ref{f10},
the branching ratios are suppressed below $10^{-12}$. After comparing the results
in Fig. \ref{f10} with the analog results of Fig. \ref{f9}, we see that the maximum
values of LFV tau decays in the case of IO are unable to exceed over that of the
corresponding decays in the case of NO, except for the decay $\tau^-\to\mu^-e^-e^+$.
Again, we make the comparison between the results of Fig. \ref{f10} and the analog
results in the MSM, which are presented in Fig. \ref{f8}. We see that, in the
case of IO, the maximal values of the branching ratios of LFV tau decays in the DSM
are unable to overcome that of the corresponding decays in the MSM.

\section{Conclusions}
\label{s7}

In this work, we have studied LFV processes in the charged lepton sector of
the two minimal versions of the scotogenic models. In one version of the scotogenic
model, neutrinos are Majorana. Hence, this model is called MSM. Whereas, in the
other version of the model, which is called DSM, neutrinos are Dirac. In both
the MSM and DSM, we have obtained the analytical expressions for the branching
ratios of all the LFV decays that are tabulated in Tab. \ref{t1}, in addition
to finding the conversion rate of $\mu{\rm Au}\to e{\rm Au}$.

After scanning
over the parameter space of the MSM and DSM, we have first studied constrains to
be imposed by the observables ${\rm Br}(\mu\to e\gamma)$, ${\rm Br}(\mu\to3e)$ and
${\rm CR}(\mu\to e,{\rm Au})$. While doing this analysis, we have emphasized the
role played by the perturbativity bound on the Yukawa couplings. We have shown
that, by satisfying the experimental limit on ${\rm Br}(\mu\to e\gamma)$ and
perturbativity bound on Yukawa couplings, there can exist a parameter region
in both the MSM and DSM, where the experimental limit on ${\rm Br}(\mu\to3e)$
cannot be satisfied. This implies that ${\rm Br}(\mu\to e\gamma)$ and
${\rm Br}(\mu\to3e)$ can give constraints of their own on the model parameters
of the MSM and DSM. We have also shown that the constraints due to ${\rm Br}(\mu\to3e)$
are stringent in the MSM in comparison to that in the DSM. This reflects to
the fact that, between these two models, additional box contribution exist in
the MSM. On the other hand,
we have not found a region in both the MSM and DSM, where
${\rm CR}(\mu\to e,{\rm Au})$ exceed over the experimental limit on this
observable, after satisfying the experimental limit on ${\rm Br}(\mu\to e\gamma)$.

Next, we have studied LFV tau decays in both the MSM and DSM, after satisfying
the experimental limits on ${\rm Br}(\mu\to e\gamma)$, ${\rm Br}(\mu\to3e)$,
${\rm CR}(\mu\to e,{\rm Au})$ and perturbativity bound on the Yukawa couplings.
In our analysis, it is found that the maximal reach of ${\rm Br}(\tau\to3\mu)$
is large compared to any other LFV tau decay in both the MSM and DSM.
The decay $\tau\to3\mu$ can have a branching ratio which can reach to around
$10^{-10}(10^{-11})$ in the MSM(DSM). The maximal reach of ${\rm Br}(\tau\to3\mu)$,
in the case of MSM, is the sensitivity region to be probed
in the future experiments. We have found that, in both
the MSM and DSM, the maximal
reaches of the branching ratios of the LFV tau decays in the case of IO cannot
exceed over that of the corresponding decays in the case of NO. We have also
compared the branching ratios of the LFV tau decays between the MSM and DSM.
In this comparison, we have found that the 2-body decays $\tau\to e\gamma$
and $\tau\to\mu\gamma$ can have same maximal values of the branching ratios
in both the MSM and DSM. On the other hand, for the 3-body LFV tau decays,
the maximal values of the branching ratios in the DSM are suppressed in
comparison to that of the corresponding branching ratios in the MSM. This is
vindicated by the fact that there exist additional contribution to the 3-body
LFV tau decays in the MSM in comparison to that in the DSM.

\bibliography{refs}

\providecommand{\href}[2]{#2}\begingroup\raggedright\begin{thebibliography}{10}

\bibitem{Kajita:2016cak}
T.~Kajita, \emph{{Nobel Lecture: Discovery of atmospheric neutrino
  oscillations}},
  \href{https://doi.org/10.1103/RevModPhys.88.030501}{\emph{Rev. Mod. Phys.}
  {\bfseries 88} (2016) 030501}.

\bibitem{McDonald:2016ixn}
A.B.~McDonald, \emph{{Nobel Lecture: The Sudbury Neutrino Observatory:
  Observation of flavor change for solar neutrinos}},
  \href{https://doi.org/10.1103/RevModPhys.88.030502}{\emph{Rev. Mod. Phys.}
  {\bfseries 88} (2016) 030502}.

\bibitem{Planck:2018vyg}
{\scshape Planck} collaboration, \emph{{Planck 2018 results. VI. Cosmological
  parameters}},
  \href{https://doi.org/10.1051/0004-6361/201833910}{\emph{Astron. Astrophys.}
  {\bfseries 641} (2020) A6}
  [\href{https://arxiv.org/abs/1807.06209}{{\ttfamily 1807.06209}}].

\bibitem{Quigg:2004is}
C.~Quigg, \emph{{Beyond the standard model in many directions}},  in \emph{{2nd
  CERN-CLAF School of High Energy Physics}}, pp.~57--118, 4, 2004
  [\href{https://arxiv.org/abs/hep-ph/0404228}{{\ttfamily hep-ph/0404228}}].

\bibitem{Ellis:2009pz}
J.~Ellis, \emph{{Physics Beyond the Standard Model}},
  \href{https://doi.org/10.1016/j.nuclphysa.2009.05.034}{\emph{Nucl. Phys. A}
  {\bfseries 827} (2009) 187C}
  [\href{https://arxiv.org/abs/0902.0357}{{\ttfamily 0902.0357}}].

\bibitem{King:2003jb}
S.F.~King, \emph{{Neutrino mass models}},
  \href{https://doi.org/10.1088/0034-4885/67/2/R01}{\emph{Rept. Prog. Phys.}
  {\bfseries 67} (2004) 107}
  [\href{https://arxiv.org/abs/hep-ph/0310204}{{\ttfamily hep-ph/0310204}}].

\bibitem{Altarelli:2006ri}
G.~Altarelli, \emph{{Models of neutrino masses and mixings}},  in \emph{{61st
  Scottish Universities Summer School in Physics: Neutrinos in Particle
  Physics, Astrophysics and Cosmology (SUSSP61)}}, pp.~91--115, 11, 2006
  [\href{https://arxiv.org/abs/hep-ph/0611117}{{\ttfamily hep-ph/0611117}}].

\bibitem{Minkowski:1977sc}
P.~Minkowski, \emph{{$\mu \to e\gamma$ at a Rate of One Out of $10^{9}$ Muon
  Decays?}}, \href{https://doi.org/10.1016/0370-2693(77)90435-X}{\emph{Phys.
  Lett. B} {\bfseries 67} (1977) 421}.

\bibitem{Gell-Mann:1979vob}
M.~Gell-Mann, P.~Ramond and R.~Slansky, \emph{{Complex Spinors and Unified
  Theories}}, {\emph{Conf. Proc. C} {\bfseries 790927} (1979) 315}
  [\href{https://arxiv.org/abs/1306.4669}{{\ttfamily 1306.4669}}].

\bibitem{Mohapatra:1979ia}
R.N.~Mohapatra and G.~Senjanovic, \emph{{Neutrino Mass and Spontaneous Parity
  Nonconservation}},
  \href{https://doi.org/10.1103/PhysRevLett.44.912}{\emph{Phys. Rev. Lett.}
  {\bfseries 44} (1980) 912}.

\bibitem{Schechter:1980gr}
J.~Schechter and J.W.F.~Valle, \emph{{Neutrino Masses in SU(2) x U(1)
  Theories}}, \href{https://doi.org/10.1103/PhysRevD.22.2227}{\emph{Phys. Rev.
  D} {\bfseries 22} (1980) 2227}.

\bibitem{Magg:1980ut}
M.~Magg and C.~Wetterich, \emph{{Neutrino Mass Problem and Gauge Hierarchy}},
  \href{https://doi.org/10.1016/0370-2693(80)90825-4}{\emph{Phys. Lett. B}
  {\bfseries 94} (1980) 61}.

\bibitem{Mohapatra:1980yp}
R.N.~Mohapatra and G.~Senjanovic, \emph{{Neutrino Masses and Mixings in Gauge
  Models with Spontaneous Parity Violation}},
  \href{https://doi.org/10.1103/PhysRevD.23.165}{\emph{Phys. Rev. D} {\bfseries
  23} (1981) 165}.

\bibitem{Lazarides:1980nt}
G.~Lazarides, Q.~Shafi and C.~Wetterich, \emph{{Proton Lifetime and Fermion
  Masses in an SO(10) Model}},
  \href{https://doi.org/10.1016/0550-3213(81)90354-0}{\emph{Nucl. Phys. B}
  {\bfseries 181} (1981) 287}.

\bibitem{Foot:1988aq}
R.~Foot, H.~Lew, X.G.~He and G.C.~Joshi, \emph{{Seesaw Neutrino Masses Induced
  by a Triplet of Leptons}}, \href{https://doi.org/10.1007/BF01415558}{\emph{Z.
  Phys. C} {\bfseries 44} (1989) 441}.

\bibitem{Mohapatra:1987hh}
R.N.~Mohapatra, \emph{{A Model for Dirac Neutrino Masses and Mixings}},
  \href{https://doi.org/10.1016/0370-2693(87)90161-4}{\emph{Phys. Lett. B}
  {\bfseries 198} (1987) 69}.

\bibitem{Chang:1986bp}
D.~Chang and R.N.~Mohapatra, \emph{{Small and Calculable Dirac Neutrino Mass}},
  \href{https://doi.org/10.1103/PhysRevLett.58.1600}{\emph{Phys. Rev. Lett.}
  {\bfseries 58} (1987) 1600}.

\bibitem{Mohapatra:1987nx}
R.N.~Mohapatra, \emph{{Left-right Symmetry and Finite One Loop Dirac Neutrino
  Mass}}, \href{https://doi.org/10.1016/0370-2693(88)90610-7}{\emph{Phys. Lett.
  B} {\bfseries 201} (1988) 517}.

\bibitem{Balakrishna:1988bn}
B.S.~Balakrishna and R.N.~Mohapatra, \emph{{Radiative Fermion Masses From New
  Physics at Tev Scale}},
  \href{https://doi.org/10.1016/0370-2693(89)91129-5}{\emph{Phys. Lett. B}
  {\bfseries 216} (1989) 349}.

\bibitem{Ma:1989tz}
E.~Ma, \emph{{Radiative Quark and Lepton Masses in a Left-right Gauge Model}},
  \href{https://doi.org/10.1103/PhysRevLett.63.1042}{\emph{Phys. Rev. Lett.}
  {\bfseries 63} (1989) 1042}.

\bibitem{Babu:1988yq}
K.S.~Babu and X.G.~He, \emph{{DIRAC NEUTRINO MASSES AS TWO LOOP RADIATIVE
  CORRECTIONS}}, \href{https://doi.org/10.1142/S0217732389000095}{\emph{Mod.
  Phys. Lett. A} {\bfseries 4} (1989) 61}.

\bibitem{Branco:1987yg}
G.C.~Branco and C.Q.~Geng, \emph{{Naturally Small Dirac Neutrino Masses in
  Superstring Theories}},
  \href{https://doi.org/10.1103/PhysRevLett.58.969}{\emph{Phys. Rev. Lett.}
  {\bfseries 58} (1987) 969}.

\bibitem{Babu:1989fg}
K.S.~Babu and E.~Ma, \emph{{Radiative Mechanisms for Generating Quark and
  Lepton Masses: Some Recent Developments}},
  \href{https://doi.org/10.1142/S0217732389002239}{\emph{Mod. Phys. Lett. A}
  {\bfseries 4} (1989) 1975}.

\bibitem{Nasri:2001ax}
S.~Nasri and S.~Moussa, \emph{{Model for small neutrino masses at the TeV
  scale}}, \href{https://doi.org/10.1142/S0217732302007119}{\emph{Mod. Phys.
  Lett. A} {\bfseries 17} (2002) 771}
  [\href{https://arxiv.org/abs/hep-ph/0106107}{{\ttfamily hep-ph/0106107}}].

\bibitem{Ma:2006km}
E.~Ma, \emph{{Verifiable radiative seesaw mechanism of neutrino mass and dark
  matter}}, \href{https://doi.org/10.1103/PhysRevD.73.077301}{\emph{Phys. Rev.
  D} {\bfseries 73} (2006) 077301}
  [\href{https://arxiv.org/abs/hep-ph/0601225}{{\ttfamily hep-ph/0601225}}].

\bibitem{Ma:2006uv}
E.~Ma, \emph{{Supersymmetric Model of Radiative Seesaw Majorana Neutrino
  Masses}}, {\emph{Annales Fond. Broglie} {\bfseries 31} (2006) 285}
  [\href{https://arxiv.org/abs/hep-ph/0607142}{{\ttfamily hep-ph/0607142}}].

\bibitem{Ma:2007yx}
E.~Ma and U.~Sarkar, \emph{{Revelations of the E(6)/U(1)(N) Model: Two-Loop
  Neutrino Mass and Dark Matter}},
  \href{https://doi.org/10.1016/j.physletb.2007.08.019}{\emph{Phys. Lett. B}
  {\bfseries 653} (2007) 288}
  [\href{https://arxiv.org/abs/0705.0074}{{\ttfamily 0705.0074}}].

\bibitem{Ma:2007gq}
E.~Ma, \emph{{Z(3) Dark Matter and Two-Loop Neutrino Mass}},
  \href{https://doi.org/10.1016/j.physletb.2008.02.053}{\emph{Phys. Lett. B}
  {\bfseries 662} (2008) 49} [\href{https://arxiv.org/abs/0708.3371}{{\ttfamily
  0708.3371}}].

\bibitem{Ma:2007kt}
E.~Ma, \emph{{SU(5) completion of the dark scalar doublet model of radiative
  neutrino mass}},
  \href{https://doi.org/10.1016/j.physletb.2007.12.013}{\emph{Phys. Lett. B}
  {\bfseries 659} (2008) 885}
  [\href{https://arxiv.org/abs/0710.2325}{{\ttfamily 0710.2325}}].

\bibitem{Ma:2008ba}
E.~Ma, \emph{{Supersymmetric U(1) Gauge Realization of the Dark Scalar Doublet
  Model of Radiative Neutrino Mass}},
  \href{https://doi.org/10.1142/S0217732308026753}{\emph{Mod. Phys. Lett. A}
  {\bfseries 23} (2008) 721} [\href{https://arxiv.org/abs/0801.2545}{{\ttfamily
  0801.2545}}].

\bibitem{Gu:2008zf}
P.-H.~Gu and U.~Sarkar, \emph{{Radiative seesaw in left-right symmetric
  model}}, \href{https://doi.org/10.1103/PhysRevD.78.073012}{\emph{Phys. Rev.
  D} {\bfseries 78} (2008) 073012}
  [\href{https://arxiv.org/abs/0807.0270}{{\ttfamily 0807.0270}}].

\bibitem{Ma:2008cu}
E.~Ma and D.~Suematsu, \emph{{Fermion Triplet Dark Matter and Radiative
  Neutrino Mass}}, \href{https://doi.org/10.1142/S021773230903059X}{\emph{Mod.
  Phys. Lett. A} {\bfseries 24} (2009) 583}
  [\href{https://arxiv.org/abs/0809.0942}{{\ttfamily 0809.0942}}].

\bibitem{Vicente:2014wga}
A.~Vicente and C.E.~Yaguna, \emph{{Probing the scotogenic model with lepton
  flavor violating processes}},
  \href{https://doi.org/10.1007/JHEP02(2015)144}{\emph{JHEP} {\bfseries 02}
  (2015) 144} [\href{https://arxiv.org/abs/1412.2545}{{\ttfamily 1412.2545}}].

\bibitem{Molinaro:2014lfa}
E.~Molinaro, C.E.~Yaguna and O.~Zapata, \emph{{FIMP realization of the
  scotogenic model}},
  \href{https://doi.org/10.1088/1475-7516/2014/07/015}{\emph{JCAP} {\bfseries
  07} (2014) 015} [\href{https://arxiv.org/abs/1405.1259}{{\ttfamily
  1405.1259}}].

\bibitem{Lindner:2016kqk}
M.~Lindner, M.~Platscher, C.E.~Yaguna and A.~Merle, \emph{{Fermionic WIMPs and
  vacuum stability in the scotogenic model}},
  \href{https://doi.org/10.1103/PhysRevD.94.115027}{\emph{Phys. Rev. D}
  {\bfseries 94} (2016) 115027}
  [\href{https://arxiv.org/abs/1608.00577}{{\ttfamily 1608.00577}}].

\bibitem{Escribano:2020iqq}
P.~Escribano, M.~Reig and A.~Vicente, \emph{{Generalizing the Scotogenic
  model}}, \href{https://doi.org/10.1007/JHEP07(2020)097}{\emph{JHEP}
  {\bfseries 07} (2020) 097}
  [\href{https://arxiv.org/abs/2004.05172}{{\ttfamily 2004.05172}}].

\bibitem{Sarazin:2021nwo}
M.~Sarazin, J.~Bernigaud and B.~Herrmann, \emph{{Dark matter and lepton flavour
  phenomenology in a singlet-doublet scotogenic model}},
  \href{https://doi.org/10.1007/JHEP12(2021)116}{\emph{JHEP} {\bfseries 12}
  (2021) 116} [\href{https://arxiv.org/abs/2107.04613}{{\ttfamily
  2107.04613}}].

\bibitem{Portillo-Sanchez:2023kbz}
D.~Portillo-S\'anchez, P.~Escribano and A.~Vicente, \emph{{Ultraviolet
  extensions of the Scotogenic model}},
  \href{https://doi.org/10.1007/JHEP08(2023)023}{\emph{JHEP} {\bfseries 08}
  (2023) 023} [\href{https://arxiv.org/abs/2301.05249}{{\ttfamily
  2301.05249}}].

\bibitem{Alvarez:2023dzz}
A.~Alvarez, A.~Banik, R.~Cepedello, B.~Herrmann, W.~Porod, M.~Sarazin et~al.,
  \emph{{Accommodating muon (g \ensuremath{-} 2) and leptogenesis in a
  scotogenic model}},
  \href{https://doi.org/10.1007/JHEP06(2023)163}{\emph{JHEP} {\bfseries 06}
  (2023) 163} [\href{https://arxiv.org/abs/2301.08485}{{\ttfamily
  2301.08485}}].

\bibitem{Abada:2023znk}
A.~Abada, N.~Bernal, A.E.~C\'arcamo~Hern\'andez, S.~Kovalenko, T.B.~de~Melo and
  T.~Toma, \emph{{Phenomenology of a scotogenic neutrino mass model at
  3-loops}},  in \emph{{18th International Conference on Topics in
  Astroparticle and Underground Physics}}, 11, 2023
  [\href{https://arxiv.org/abs/2311.14716}{{\ttfamily 2311.14716}}].

\bibitem{Ahriche:2023hho}
A.~Ahriche, M.L.~Bellilet, M.O.~Khojali, M.~Kumar and A.-T.~Mulaudzi,
  \emph{{The scale invariant scotogenic model: CDF-II $W$-boson mass and the
  95\textasciitilde{}GeV excesses}},
  \href{https://arxiv.org/abs/2311.08297}{{\ttfamily 2311.08297}}.

\bibitem{Borah:2023hqw}
D.~Borah, S.~Mahapatra, P.K.~Paul and N.~Sahu, \emph{{Scotogenic
  $U(1)_{L_{\mu}-L_{\tau}}$ origin of $(g-2)_\mu$, W-mass anomaly and 95 GeV
  excess}},  \href{https://arxiv.org/abs/2310.11953}{{\ttfamily 2310.11953}}.

\bibitem{Avnish:2023wxk}
Avnish, \emph{{Maximum Yukawa Couplings for WIMP Majorana Dark Matter in
  Scotogenic Model}},
  \href{https://doi.org/10.1134/S1547477123050084}{\emph{Phys. Part. Nucl.
  Lett.} {\bfseries 20} (2023) 1146}.

\bibitem{Singh:2023eye}
L.~Singh, D.~Mahanta and S.~Verma, \emph{{Low Scale Leptogenesis in
  Singlet-Triplet Scotogenic Model}},
  \href{https://arxiv.org/abs/2309.12755}{{\ttfamily 2309.12755}}.

\bibitem{Kitabayashi:2023rje}
T.~Kitabayashi, \emph{{Constraining scotogenic dark matter and primordial black
  holes using gravitational waves}},
  \href{https://arxiv.org/abs/2309.01883}{{\ttfamily 2309.01883}}.

\bibitem{Karan:2023adm}
A.~Karan, S.~Sadhukhan and J.W.F.~Valle, \emph{{Phenomenological profile of
  scotogenic fermionic dark matter}},
  \href{https://arxiv.org/abs/2308.09135}{{\ttfamily 2308.09135}}.

\bibitem{Herms:2023cyy}
J.~Herms, S.~Jana, V.P.~K. and S.~Saad, \emph{{Light neutrinophilic dark matter
  from a scotogenic model}},
  \href{https://doi.org/10.1016/j.physletb.2023.138167}{\emph{Phys. Lett. B}
  {\bfseries 845} (2023) 138167}
  [\href{https://arxiv.org/abs/2307.15760}{{\ttfamily 2307.15760}}].

\bibitem{Ismael:2023jlp}
M.S.~Ismael, G.~Faisel and M.~Alanssari, \emph{{The Impact of Dark Matter on
  the Related Sector of the Scotogenic Model and Its Implications}},
  \href{https://doi.org/10.31526/lhep.2023.374}{\emph{LHEP} {\bfseries 2023}
  (2023) 374}.

\bibitem{Escribano:2023hxj}
P.~Escribano, V.M.~Lozano and A.~Vicente, \emph{{A Scotogenic explanation for
  the 95 GeV excesses}},  \href{https://arxiv.org/abs/2306.03735}{{\ttfamily
  2306.03735}}.

\bibitem{Hundi:2022iva}
R.S.~Hundi, \emph{{Lepton flavor violating Z and Higgs decays in the scotogenic
  model}}, \href{https://doi.org/10.1140/epjc/s10052-022-10453-3}{\emph{Eur.
  Phys. J. C} {\bfseries 82} (2022) 505}
  [\href{https://arxiv.org/abs/2201.03779}{{\ttfamily 2201.03779}}].

\bibitem{Hundi:2015kea}
R.S.~Hundi, \emph{{\ensuremath{\mu}\textrightarrow{}e\ensuremath{\gamma} in a
  supersymmetric radiative neutrino mass model}},
  \href{https://doi.org/10.1103/PhysRevD.93.015008}{\emph{Phys. Rev. D}
  {\bfseries 93} (2016) 015008}
  [\href{https://arxiv.org/abs/1510.02253}{{\ttfamily 1510.02253}}].

\bibitem{Farzan:2012sa}
Y.~Farzan and E.~Ma, \emph{{Dirac neutrino mass generation from dark matter}},
  \href{https://doi.org/10.1103/PhysRevD.86.033007}{\emph{Phys. Rev. D}
  {\bfseries 86} (2012) 033007}
  [\href{https://arxiv.org/abs/1204.4890}{{\ttfamily 1204.4890}}].

\bibitem{Ma:2016mwh}
E.~Ma and O.~Popov, \emph{{Pathways to Naturally Small Dirac Neutrino Masses}},
  \href{https://doi.org/10.1016/j.physletb.2016.11.027}{\emph{Phys. Lett. B}
  {\bfseries 764} (2017) 142}
  [\href{https://arxiv.org/abs/1609.02538}{{\ttfamily 1609.02538}}].

\bibitem{Wang:2017mcy}
W.~Wang, R.~Wang, Z.-L.~Han and J.-Z.~Han, \emph{{The $B-L$ Scotogenic Models
  for Dirac Neutrino Masses}},
  \href{https://doi.org/10.1140/epjc/s10052-017-5446-9}{\emph{Eur. Phys. J. C}
  {\bfseries 77} (2017) 889}
  [\href{https://arxiv.org/abs/1705.00414}{{\ttfamily 1705.00414}}].

\bibitem{Yao:2017vtm}
C.-Y.~Yao and G.-J.~Ding, \emph{{Systematic Study of One-Loop Dirac Neutrino
  Masses and Viable Dark Matter Candidates}},
  \href{https://doi.org/10.1103/PhysRevD.96.095004}{\emph{Phys. Rev. D}
  {\bfseries 96} (2017) 095004}
  [\href{https://arxiv.org/abs/1707.09786}{{\ttfamily 1707.09786}}].

\bibitem{Calle:2018ovc}
J.~Calle, D.~Restrepo, C.E.~Yaguna and O.~Zapata, \emph{{Minimal radiative
  Dirac neutrino mass models}},
  \href{https://doi.org/10.1103/PhysRevD.99.075008}{\emph{Phys. Rev. D}
  {\bfseries 99} (2019) 075008}
  [\href{https://arxiv.org/abs/1812.05523}{{\ttfamily 1812.05523}}].

\bibitem{Ma:2019yfo}
E.~Ma, \emph{{Scotogenic $U(1)_\chi$ Dirac neutrinos}},
  \href{https://doi.org/10.1016/j.physletb.2019.05.006}{\emph{Phys. Lett. B}
  {\bfseries 793} (2019) 411}
  [\href{https://arxiv.org/abs/1901.09091}{{\ttfamily 1901.09091}}].

\bibitem{Ma:2019iwj}
E.~Ma, \emph{{Scotogenic cobimaximal Dirac neutrino mixing from $\Delta (27)$
  and $U(1)_\chi $}},
  \href{https://doi.org/10.1140/epjc/s10052-019-7440-x}{\emph{Eur. Phys. J. C}
  {\bfseries 79} (2019) 903}
  [\href{https://arxiv.org/abs/1905.01535}{{\ttfamily 1905.01535}}].

\bibitem{Ma:2019byo}
E.~Ma, \emph{{Two-loop $Z_4$ Dirac neutrino masses and mixing, with
  self-interacting dark matter}},
  \href{https://doi.org/10.1016/j.nuclphysb.2019.114725}{\emph{Nucl. Phys. B}
  {\bfseries 946} (2019) 114725}
  [\href{https://arxiv.org/abs/1907.04665}{{\ttfamily 1907.04665}}].

\bibitem{CentellesChulia:2019xky}
S.~Centelles~Chuli\'a, R.~Cepedello, E.~Peinado and R.~Srivastava,
  \emph{{Systematic classification of two loop $d$ = 4 Dirac neutrino mass
  models and the Diracness-dark matter stability connection}},
  \href{https://doi.org/10.1007/JHEP10(2019)093}{\emph{JHEP} {\bfseries 10}
  (2019) 093} [\href{https://arxiv.org/abs/1907.08630}{{\ttfamily
  1907.08630}}].

\bibitem{Jana:2019mgj}
S.~Jana, P.K.~Vishnu and S.~Saad, \emph{{Minimal realizations of Dirac neutrino
  mass from generic one-loop and two-loop topologies at $d = 5$}},
  \href{https://doi.org/10.1088/1475-7516/2020/04/018}{\emph{JCAP} {\bfseries
  04} (2020) 018} [\href{https://arxiv.org/abs/1910.09537}{{\ttfamily
  1910.09537}}].

\bibitem{Leite:2020wjl}
J.~Leite, A.~Morales, J.W.F.~Valle and C.A.~Vaquera-Araujo, \emph{{Scotogenic
  dark matter and Dirac neutrinos from unbroken gauged B \ensuremath{-} L
  symmetry}}, \href{https://doi.org/10.1016/j.physletb.2020.135537}{\emph{Phys.
  Lett. B} {\bfseries 807} (2020) 135537}
  [\href{https://arxiv.org/abs/2003.02950}{{\ttfamily 2003.02950}}].

\bibitem{Guo:2020qin}
S.-Y.~Guo and Z.-L.~Han, \emph{{Observable Signatures of Scotogenic Dirac
  Model}}, \href{https://doi.org/10.1007/JHEP12(2020)062}{\emph{JHEP}
  {\bfseries 12} (2020) 062}
  [\href{https://arxiv.org/abs/2005.08287}{{\ttfamily 2005.08287}}].

\bibitem{Bernal:2021ezl}
N.~Bernal, J.~Calle and D.~Restrepo, \emph{{Anomaly-free Abelian gauge
  symmetries with Dirac scotogenic models}},
  \href{https://doi.org/10.1103/PhysRevD.103.095032}{\emph{Phys. Rev. D}
  {\bfseries 103} (2021) 095032}
  [\href{https://arxiv.org/abs/2102.06211}{{\ttfamily 2102.06211}}].

\bibitem{De:2021crr}
B.~De, D.~Das, M.~Mitra and N.~Sahoo, \emph{{Magnetic moments of leptons,
  charged lepton flavor violations and dark matter phenomenology of a minimal
  radiative Dirac neutrino mass model}},
  \href{https://doi.org/10.1007/JHEP08(2022)202}{\emph{JHEP} {\bfseries 08}
  (2022) 202} [\href{https://arxiv.org/abs/2106.00979}{{\ttfamily
  2106.00979}}].

\bibitem{Mishra:2021ilq}
S.~Mishra, N.~Narendra, P.K.~Panda and N.~Sahoo, \emph{{Scalar dark matter and
  radiative Dirac neutrino mass in an extended U(1)B\ensuremath{-}L model}},
  \href{https://doi.org/10.1016/j.nuclphysb.2022.115855}{\emph{Nucl. Phys. B}
  {\bfseries 981} (2022) 115855}
  [\href{https://arxiv.org/abs/2112.12569}{{\ttfamily 2112.12569}}].

\bibitem{Chowdhury:2022jde}
T.A.~Chowdhury, M.~Ehsanuzzaman and S.~Saad, \emph{{Dark Matter and (g -
  2)$_{\mu,e}$ in radiative Dirac neutrino mass models}},
  \href{https://doi.org/10.1088/1475-7516/2022/08/076}{\emph{JCAP} {\bfseries
  08} (2022) 076} [\href{https://arxiv.org/abs/2203.14983}{{\ttfamily
  2203.14983}}].

\bibitem{Li:2022chc}
S.-P.~Li, X.-Q.~Li, X.-S.~Yan and Y.-D.~Yang, \emph{{Scotogenic Dirac neutrino
  mass models embedded with leptoquarks: one pathway to address the flavor
  anomalies and the neutrino masses together}},
  \href{https://doi.org/10.1140/epjc/s10052-022-11054-w}{\emph{Eur. Phys. J. C}
  {\bfseries 82} (2022) 1078}
  [\href{https://arxiv.org/abs/2204.09201}{{\ttfamily 2204.09201}}].

\bibitem{Hazarika:2022tlc}
N.~Hazarika and K.~Bora, \emph{{A new viable mass region of dark matter and
  dirac neutrino mass generation in a scotogenic extension of SM}},
  \href{https://doi.org/10.1142/S0217751X23500513}{\emph{Int. J. Mod. Phys. A}
  {\bfseries 38} (2023) 2350051}
  [\href{https://arxiv.org/abs/2205.06003}{{\ttfamily 2205.06003}}].

\bibitem{Borah:2022enh}
D.~Borah, P.~Das and D.~Nanda, \emph{{Observable ${\rm \Delta{N_{eff}}}$ in
  Dirac Scotogenic Model}},  \href{https://arxiv.org/abs/2211.13168}{{\ttfamily
  2211.13168}}.

\bibitem{Hundi:2023tdq}
R.S.~Hundi, \emph{{Study on the global minimum and
  H\textrightarrow{}\ensuremath{\gamma}\ensuremath{\gamma} in the Dirac
  scotogenic model}},
  \href{https://doi.org/10.1103/PhysRevD.108.015006}{\emph{Phys. Rev. D}
  {\bfseries 108} (2023) 015006}
  [\href{https://arxiv.org/abs/2303.04655}{{\ttfamily 2303.04655}}].

\bibitem{Rojas:2018wym}
N.~Rojas, R.~Srivastava and J.W.F.~Valle, \emph{{Simplest Scoto-Seesaw
  Mechanism}},
  \href{https://doi.org/10.1016/j.physletb.2018.12.014}{\emph{Phys. Lett. B}
  {\bfseries 789} (2019) 132}
  [\href{https://arxiv.org/abs/1807.11447}{{\ttfamily 1807.11447}}].

\bibitem{Kumar:2023moh}
R.~Kumar, P.~Mishra, M.K.~Behera, R.~Mohanta and R.~Srivastava,
  \emph{{Predictions from scoto-seesaw with A4 modular symmetry}},
  \href{https://doi.org/10.1016/j.physletb.2024.138635}{\emph{Phys. Lett. B}
  {\bfseries 853} (2024) 138635}
  [\href{https://arxiv.org/abs/2310.02363}{{\ttfamily 2310.02363}}].

\bibitem{Kumar:2024zfb}
R.~Kumar, N.~Nath and R.~Srivastava, \emph{{Cutting the scotogenic loop: adding
  flavor to dark matter}},
  \href{https://doi.org/10.1007/JHEP12(2024)036}{\emph{JHEP} {\bfseries 12}
  (2024) 036} [\href{https://arxiv.org/abs/2406.00188}{{\ttfamily
  2406.00188}}].

\bibitem{deSalas:2020pgw}
P.F.~de~Salas, D.V.~Forero, S.~Gariazzo, P.~Mart\'\i{}nez-Mirav\'e, O.~Mena,
  C.A.~Ternes et~al., \emph{{2020 global reassessment of the neutrino
  oscillation picture}},
  \href{https://doi.org/10.1007/JHEP02(2021)071}{\emph{JHEP} {\bfseries 02}
  (2021) 071} [\href{https://arxiv.org/abs/2006.11237}{{\ttfamily
  2006.11237}}].

\bibitem{MEGII:2023ltw}
{\scshape MEG II} collaboration, \emph{{A search for $\mu ^+ \rightarrow
  \mathrm{e}^+ \gamma $ with the first dataset of the MEG~II experiment}},
  \href{https://doi.org/10.1140/epjc/s10052-024-12416-2}{\emph{Eur. Phys. J. C}
  {\bfseries 84} (2024) 216}
  [\href{https://arxiv.org/abs/2310.12614}{{\ttfamily 2310.12614}}].

\bibitem{BaBar:2009hkt}
{\scshape BaBar} collaboration, \emph{{Searches for Lepton Flavor Violation in
  the Decays tau+- ---\ensuremath{>} e+- gamma and tau+- ---\ensuremath{>} mu+-
  gamma}}, \href{https://doi.org/10.1103/PhysRevLett.104.021802}{\emph{Phys.
  Rev. Lett.} {\bfseries 104} (2010) 021802}
  [\href{https://arxiv.org/abs/0908.2381}{{\ttfamily 0908.2381}}].

\bibitem{Belle:2021ysv}
{\scshape Belle} collaboration, \emph{{Search for lepton-flavor-violating
  tau-lepton decays to $\ell\gamma$ at Belle}},
  \href{https://doi.org/10.1007/JHEP10(2021)019}{\emph{JHEP} {\bfseries 10}
  (2021) 19} [\href{https://arxiv.org/abs/2103.12994}{{\ttfamily 2103.12994}}].

\bibitem{SINDRUM:1987nra}
{\scshape SINDRUM} collaboration, \emph{{Search for the Decay $\mu^+ \to e^+
  e^+ e^-$}}, \href{https://doi.org/10.1016/0550-3213(88)90462-2}{\emph{Nucl.
  Phys. B} {\bfseries 299} (1988) 1}.

\bibitem{Hayasaka:2010np}
K.~Hayasaka et~al., \emph{{Search for Lepton Flavor Violating Tau Decays into
  Three Leptons with 719 Million Produced Tau+Tau- Pairs}},
  \href{https://doi.org/10.1016/j.physletb.2010.03.037}{\emph{Phys. Lett. B}
  {\bfseries 687} (2010) 139}
  [\href{https://arxiv.org/abs/1001.3221}{{\ttfamily 1001.3221}}].

\bibitem{SINDRUMII:2006dvw}
{\scshape SINDRUM II} collaboration, \emph{{A Search for muon to electron
  conversion in muonic gold}},
  \href{https://doi.org/10.1140/epjc/s2006-02582-x}{\emph{Eur. Phys. J. C}
  {\bfseries 47} (2006) 337}.

\bibitem{SINDRUMII:1993gxf}
{\scshape SINDRUM II} collaboration, \emph{{Test of lepton flavor conservation
  in mu ---\ensuremath{>} e conversion on titanium}},
  \href{https://doi.org/10.1016/0370-2693(93)91383-X}{\emph{Phys. Lett. B}
  {\bfseries 317} (1993) 631}.

\bibitem{Toma:2013zsa}
T.~Toma and A.~Vicente, \emph{{Lepton Flavor Violation in the Scotogenic
  Model}}, \href{https://doi.org/10.1007/JHEP01(2014)160}{\emph{JHEP}
  {\bfseries 01} (2014) 160} [\href{https://arxiv.org/abs/1312.2840}{{\ttfamily
  1312.2840}}].

\bibitem{Workman:2022ynf}
{\scshape Particle Data Group} collaboration, \emph{{Review of Particle
  Physics}}, \href{https://doi.org/10.1093/ptep/ptac097}{\emph{PTEP} {\bfseries
  2022} (2022) 083C01}.

\bibitem{Kubo:2006yx}
J.~Kubo, E.~Ma and D.~Suematsu, \emph{{Cold Dark Matter, Radiative Neutrino
  Mass, $\mu \to e\gamma$, and Neutrinoless Double Beta Decay}},
  \href{https://doi.org/10.1016/j.physletb.2006.08.085}{\emph{Phys. Lett. B}
  {\bfseries 642} (2006) 18}
  [\href{https://arxiv.org/abs/hep-ph/0604114}{{\ttfamily hep-ph/0604114}}].

\bibitem{Kitabayashi:2021fqx}
T.~Kitabayashi, \emph{{Primordial black holes and lepton flavor violation with
  scotogenic dark matter}},
  \href{https://doi.org/10.1093/ptep/ptac025}{\emph{PTEP} {\bfseries 2022}
  (2022) 033B02} [\href{https://arxiv.org/abs/2107.11692}{{\ttfamily
  2107.11692}}].

\bibitem{DeRomeri:2022cem}
V.~De~Romeri, J.~Nava, M.~Puerta and A.~Vicente, \emph{{Dark matter in the
  scotogenic model with spontaneous lepton number violation}},
  \href{https://doi.org/10.1103/PhysRevD.107.095019}{\emph{Phys. Rev. D}
  {\bfseries 107} (2023) 095019}
  [\href{https://arxiv.org/abs/2210.07706}{{\ttfamily 2210.07706}}].

\bibitem{Tapender:2024ktc}
Tapender, S.~Verma and S.~Kumar, \emph{{On lepton flavor violation and dark
  matter in Scotogenic model with trimaximal mixing}},
  \href{https://doi.org/10.1140/epjp/s13360-024-05950-1}{\emph{Eur. Phys. J.
  Plus} {\bfseries 140} (2025) 43}
  [\href{https://arxiv.org/abs/2402.16491}{{\ttfamily 2402.16491}}].

\bibitem{Casas:2001sr}
J.A.~Casas and A.~Ibarra, \emph{{Oscillating neutrinos and $\mu \to e,
  \gamma$}}, \href{https://doi.org/10.1016/S0550-3213(01)00475-8}{\emph{Nucl.
  Phys. B} {\bfseries 618} (2001) 171}
  [\href{https://arxiv.org/abs/hep-ph/0103065}{{\ttfamily hep-ph/0103065}}].

\bibitem{Grimus:2002ux}
W.~Grimus and L.~Lavoura, \emph{{Soft lepton flavor violation in a multi Higgs
  doublet seesaw model}},
  \href{https://doi.org/10.1103/PhysRevD.66.014016}{\emph{Phys. Rev. D}
  {\bfseries 66} (2002) 014016}
  [\href{https://arxiv.org/abs/hep-ph/0204070}{{\ttfamily hep-ph/0204070}}].

\bibitem{Kuno:1999jp}
Y.~Kuno and Y.~Okada, \emph{{Muon decay and physics beyond the standard
  model}}, \href{https://doi.org/10.1103/RevModPhys.73.151}{\emph{Rev. Mod.
  Phys.} {\bfseries 73} (2001) 151}
  [\href{https://arxiv.org/abs/hep-ph/9909265}{{\ttfamily hep-ph/9909265}}].

\bibitem{Chiang:1993xz}
H.C.~Chiang, E.~Oset, T.S.~Kosmas, A.~Faessler and J.D.~Vergados,
  \emph{{Coherent and incoherent (mu-, e-) conversion in nuclei}},
  \href{https://doi.org/10.1016/0375-9474(93)90259-Z}{\emph{Nucl. Phys. A}
  {\bfseries 559} (1993) 526}.

\bibitem{Kosmas:2001mv}
T.S.~Kosmas, S.~Kovalenko and I.~Schmidt, \emph{{Nuclear muon- e- conversion in
  strange quark sea}},
  \href{https://doi.org/10.1016/S0370-2693(01)00657-8}{\emph{Phys. Lett. B}
  {\bfseries 511} (2001) 203}
  [\href{https://arxiv.org/abs/hep-ph/0102101}{{\ttfamily hep-ph/0102101}}].

\bibitem{Kitano:2002mt}
R.~Kitano, M.~Koike and Y.~Okada, \emph{{Detailed calculation of lepton flavor
  violating muon electron conversion rate for various nuclei}},
  \href{https://doi.org/10.1103/PhysRevD.76.059902}{\emph{Phys. Rev. D}
  {\bfseries 66} (2002) 096002}
  [\href{https://arxiv.org/abs/hep-ph/0203110}{{\ttfamily hep-ph/0203110}}].

\bibitem{Arganda:2007jw}
E.~Arganda, M.J.~Herrero and A.M.~Teixeira, \emph{{mu-e conversion in nuclei
  within the CMSSM seesaw: Universality versus non-universality}},
  \href{https://doi.org/10.1088/1126-6708/2007/10/104}{\emph{JHEP} {\bfseries
  10} (2007) 104} [\href{https://arxiv.org/abs/0707.2955}{{\ttfamily
  0707.2955}}].

\bibitem{Belle-II:2018jsg}
{\scshape Belle-II} collaboration, \emph{{The Belle II Physics Book}},
  \href{https://doi.org/10.1093/ptep/ptz106}{\emph{PTEP} {\bfseries 2019}
  (2019) 123C01} [\href{https://arxiv.org/abs/1808.10567}{{\ttfamily
  1808.10567}}].

\bibitem{FCC:2018byv}
{\scshape FCC} collaboration, \emph{{FCC Physics Opportunities}: {Future
  Circular Collider Conceptual Design Report Volume 1}},
  \href{https://doi.org/10.1140/epjc/s10052-019-6904-3}{\emph{Eur. Phys. J. C}
  {\bfseries 79} (2019) 474}.

\end{thebibliography}\endgroup
\bibliographystyle{JHEP}
\end{document}